\documentclass{iopart}
\usepackage[latin1]{inputenc}
\usepackage{iopams}
\usepackage{color}
\usepackage[dvips]{graphicx}
\usepackage{subfigure}
\usepackage{appendix}
\begin{document}

\title[ Phase-space dynamics of a wave packet in strong fields]{Quantum and semiclassical phase-space dynamics of a wave packet in strong fields using initial-value representations}
\author{C. Zagoya$^1$, J. Wu$^1$, M. Ronto$^2$, D. V. Shalashilin$^2$ and C. Figueira de Morisson Faria$^1$}
\address{$^1$Department of Physics and Astronomy, University College London, Gower
Street, London WC1E 6BT, United Kingdom}
\address{$^2$School of Chemistry, University of Leeds, Leeds LS2 9JT, United Kingdom}
\date{\today}

\begin{abstract}
We assess the suitability of quantum and semiclassical initial value representations, exemplified by the
coupled coherent states (CCS) method and the Herman Kluk (HK) propagator, respectively, for
modeling the dynamics of an electronic wave packet in a strong laser field,
if this wave packet is initially bound. Using Wigner quasiprobability distributions and ensembles of classical trajectories,
we identify signatures of over-the-barrier and tunnel ionization in phase space for static and time-dependent fields 
and the relevant sets of phase-space trajectories in order to model such features.  Overall, we find good agreement 
with the full solution of the time-dependent Schr\"odinger equation (TDSE) for Wigner distributions constructed
with both initial-value representations. 
Our results indicate that the HK propagator does not fully account for tunneling and over-the-barrier reflections.
However, it is able to partly reproduce features associated with the wave packet crossing classically forbidden 
regions, although the trajectories employed in its construction always obey classical phase-space constraints. 
We also show that the Coupled Coherent States (CCS) method represents a fully quantum initial value representation 
and accurately reproduces the results of a standard TDSE solver. Furthermore, we sow that both the HK propagator and the CCS approach may be successfully employed to compute the time-dependent dipole acceleration and high-harmonic spectra. Nevertheless, the semiclassical propagator exhibits a worse agreement with the TDSE than the outcome of the CCS method, as it neither fully accounts for tunneling nor for over-the-barrier reflections. This leads to a dephasing in the time-dependent wave function which becomes more pronounced for longer times.
\end{abstract}

\maketitle

\section{Introduction}

Initial-value representations (IVRs) 
such as the coupled coherent states method \cite{Shalashilin_2004} and the Herman Kluk propagator \cite{Herman_1984} are widely used in many areas of science. These approaches allow an intuitive interpretation of a time-dependent wave packet in terms of trajectories in phase space, and account for
binding potentials, external fields and quantum-interference effects. Furthermore, the numerical effort in IVRs does not scale exponentially with the degrees of freedom involved. This efficiency may be increased by employing several strategies, such as dominant Hamiltonians in specific phase-space regions \cite{Zagoya_2012,Zagoya_2012_2}, or quantum-state reprojection \cite{Shalashilin_Jackson_2000,Burant_Batista_2002}.

In principle, all these features make initial-value representations very
attractive to strong-field and attosecond science. In fact, it is well known
since two decades that strong-field phenomena such as high-order harmonic
generation (HHG), above-threshold ionization (ATI) or nonsequential double
ionization (NSDI), may be described as the result of the laser-induced
scattering or recombination of an electron with its parent ion \cite{Corkum_1993}. This implies
that electron orbits play a very important role in the understanding of
these phenomena.  This has led not only to a myriad of applications, such as the attosecond
imaging of dynamic processes in matter (see, e.g., \cite{Lein_Review,Krausz_Review,Haessler_Review}), but to the extensive use of orbit-based approaches (see, e.g., \cite{Salieres_2001}).

 In particular, classical and semiclassical methods are very
popular. Classically, ensembles of electrons that behave according to the
above-mentioned recollision picture are constructed in order to mimic the behavior of the
quantum mechanical wave packet and both the external laser field and the
binding potentials are fully incorporated. This is the key idea behind
classical-trajectory methods, which have  reproduced key features such as the low-energy
structure in ATI \cite{Quan_2009} and the V-shaped structure observed in NSDI \cite{Emmanouilidou_2008,Ye_2008_2}. These methods, however, cannot account for quantum
interference effects, or tunnel ionization. Most quantum-mechanical and
semi-classical methods in strong-field physics, on the other hand,
incorporate such effects, but make drastic approximations on the residual
binding potential. A typical example is the strong-field approximation
(SFA), which, in conjunction with the steepest descent method, is the most
used approach in this field. In the SFA, the continuum is approximated by
field-dressed plane waves, i.e., the potential is not accounted for in the
electron propagation. Nonetheless, the interplay between the external laser field and the binding potential is important and has revealed itself in many ways. For instance,  this interplay leads to a prominent low-frequency structure \cite{Blaga_2009,Quan_2009,Tian-Min_2010,Telnov_2011} and
fan-shaped interference patterns \cite{Rudenko_2004,Arbo_2006} in ATI, and strongly influences NSDI in circularly polarized fields \cite{Mauger_2010,Kamor_2013}.
Because this interplay is important, in the past few years, Coulomb-corrected analytic
approaches have been developed and successfully applied to strong-field phenomena \cite{Popruzhenko_2008,Tian-Min_2010,Smirnova_2006,Smirnova_2008}. These approaches, however, require the
external field to be dominant. 

On the other hand, initial-value representations have only been employed in strong-field physics in relatively few publications.  Specifically, in \cite{Sand_1999,Zagoya_2012,Zagoya_2012_2}  HHG spectra have been computed using the Herman Kluk (HK) propagator, and in \cite{Shalashilin_2007,Kirrander_2011,Guo_2010}  NSDI ion and electron momentum distributions have been calculated employing the Coupled Coherent States (CCS) method. In order to be able to apply these methods widely, major challenges must be overcome. Concretely, the trajectories
employed in the construction of the time-dependent wave function with the HK
propagator are real, i.e., they cannot cross classically forbidden regions
in phase space. Hence, tunnel ionization may not be properly
accounted for. Since the 1990s, there has been considerable debate about whether one may model tunneling employing semiclassical IVRs such as the HK propagator, and if so, to which extent (see, e.g., \cite{Keshavamurthy_1994,Grossmann_1995,Maitra_1997,Kay_1997}). In
order to circumvent this problem, in \cite{Sand_1999,Zagoya_2012,Zagoya_2012_2} the initial electronic wave packet has
been placed far away from the core. Unfortunately, these initial conditions
leave out many strong-field problems, for which tunneling is expected to be the dominant ionization mechanism. On the 
other hand, tunneling is present in the CCS method, because it is a basis-set method. 
However, special effort must be made to choose a trajectory-guided basis which is suitable for tunneling.

Nevertheless, one may wonder whether the over-the-barrier dynamics, per se,
would not be sufficient for the modeling of strong field wave-packet
dynamics in the presence of the Coulomb potential. This is a legitimate
question, especially if one considers that classical models, for
which tunnel ionization does not occur, have been hugely successful in
reproducing a number of features in ATI and NSDI. In some of these methods,
tunnel ionization has been mimicked by employing the quasi-static
Ammosov-Delone-Krainov (ADK) tunneling rate, which may explain this success. However, there exist also purely
classical models in which the electron ensemble is left to propagate
without the need for any ad-hoc quantum mechanical ingredient. For a discussion of these models in the NSDI context see our review article \cite{FL_2011}.

For that reason, in this work we perform a systematic analysis of the
dynamics of an electronic wave packet in a strong field, with particular
focus on ionization, and on what is left out by initial value
representations. This analysis
will be performed in phase space, for reduced
dimensionality models, under the assumption that the electronic wave packet
is initially bound. As a benchmark, we will employ the full solution of the time-dependent Schr\"odinger equation (TDSE).

This article is organized as follows. In Sec.~\ref{model}, we provide the
necessary theoretical background in order to understand our
results. Subsequently, in Sec.~\ref{phasespace}, we will have a closer look at how the
time-dependent wave packet overcomes the potential barrier resulting from
the combined action of the external field and the binding potential. For
that purpose, we will employ quasiprobability distributions in phase space,
for a static and a time dependent field, and for long- and short-range potentials. As, for
short-range potentials, tunneling is expected to be more prevalent than
over-the-barrier ionization, in this case we will also perform a comparison
with the CCS method. In Sec.~\ref{estimates}, we present approximate estimates, in which the transmission through the potential barrier is computed analytically using uniform WKB approximations, and a comparison with an inverted harmonic oscillator is made. In Sec.~\ref{HHG}, we will show how the phase differences between the semiclassical and quantum mechanical computations lead to discrepancies in HHG spectra and the time-dependent dipole acceleration. In Sec.~\ref{conclusions}, we will provide a
summary of the main results and the conclusions to be drawn from this work. Finally, in the Appendix we discuss the fact that the HK and the CCS methods share a common origin.

\section{Background}
\label{model}
\subsection{Model}
For the sake of simplicity, we consider a one-electron, one-dimensional atom. The Hamiltonian of such a system can
be expressed as
\begin{equation}
\hat{H}=\frac{\hat{p}^2}{2}+\hat{V}_a+\hat{V}_{\mathcal{E}},\label{eq:Hoperator}
\end{equation}
where $\hat{V}_a$ and $\hat{V}_{\mathcal{E}}$ represents the binding potential and the interaction with the laser field in the length gauge, respectively, and the hats denote operators. Atomic units are used throughout.

The external field exhibits the same temporal dependence as in \cite{Sand_1999,Zagoya_2012}, i.e.,
\begin{equation}
\hat{V}_{\mathcal{E}}(\hat{x},t)=\mathcal{E}\hat{x}\cos(\omega t),  \label{eq:laserpot}
\label{eq:TDPotential}
\end{equation}
with $\mathcal{E}$ and $\omega$ the intensity and frequency, respectively,
of the laser field.  The binding potential is taken either as the soft-core potential
\begin{equation}
\hat{V}_{\mathrm{sc}}(\hat{x})=-\frac{1}{\sqrt{\hat{x}^2+\lambda}},  \label{eq:softcore}
\end{equation}
with $\lambda=1$ constant, or as the short-range Gaussian
potential
\begin{equation}
\hat{V}_{G}(x)=-\exp(-\lambda \hat{x}^2), \label{eq:gaussian}
\end{equation}
where $\lambda=1/2$. 
As a benchmark, we employ the full solution of the time-dependent
Schr\"odinger equation
\begin{equation}
\mathrm{i}\partial_t |\Psi(t)\rangle=\hat{H}|\Psi(t)\rangle, \label{eq:tdsegeneral}
\end{equation}
which is solved in coordinate space for $\Psi(x,t)=\langle x|\Psi(t)\rangle$ using a split operator method.

The initial wave function is defined as the Gaussian wave packet
\begin{equation}  \label{eq:initialwp}
\langle x|\Psi(0)\rangle=\left(\frac{\gamma}{\pi}\right)^{1/4}\exp\bigg\{-%
\frac{\gamma}{2}(x-q_{\alpha})^2+\mathrm{i}p_{\alpha}(x-q_{\alpha})\bigg\}\,
\end{equation}
centered at $p_{\alpha},q_{\alpha}$, with $q_{\alpha}=0$. Furthermore, the parameter $\gamma$ defines the width
of the wave packet. This initial choice facilitates the implementation of initial-value representations.

The initial energy
expectation value, computed for the system in the absence of the external driving field, is given as
\begin{equation}
\langle E(t=0)\rangle=\frac{\gamma}{4}+\frac{p_{\alpha}^2}{2}+\langle\Psi(0)|\hat{V}_a(\hat{x})|\Psi(0)\rangle.
\label{eq:energy}
\end{equation}
If the wavepacket is initially centered at $q_{\alpha}=0$, this expectation value reads as
\begin{equation}
\langle\Psi(0)|\hat{V}_G(\hat{x})|\Psi(0)\rangle=-\sqrt{\frac{\gamma  }{\gamma
   +\lambda}}
\end{equation}
and
\begin{equation}
\langle\Psi(0)|\hat{V}_{\mathrm{sc}}(\hat{x})|\Psi(0)\rangle=-\left(\frac{\gamma}{\pi}%
\right)^{1/2}K_0\left(\frac{\gamma}{2}\lambda\right)\exp\left(\frac{\gamma}{2}%
\lambda\right)\,
\end{equation} for the Gaussian and the soft-core potentials, respectively,
where $K_n(z)$ is the modified Bessel function of the second kind. Eq.~(\ref{eq:energy}) depends on
the width and the initial momentum of the initial wave packet, and on the potential parameter $\lambda$.

The width and the position of the initial wave packet were chosen to minimize its initial energy. This procedure yields $(p_{\alpha},q_{\alpha})=(0,0)$ and $\gamma\simeq0.46$ and $(p_{\alpha},q_{\alpha})=(0,0)$ and
$\gamma\simeq0.65$ for the soft-core and the Gaussian potentials, respectively. The ground-state energies associated with the potentials $V_{\mathrm{sc}}(x)$ and
$V_{G}(x)$ are $E_{\mathrm{sc}}=-0.67$ a.u. and $E_{\mathrm{G}}=-0.594$, respectively.

\subsection{Quantum and semiclassical initial value representations}
Although quantum mechanics is usually in coordinate or momentum space, it can also be formulated in phase space. In this section we %provide working equations for 
present the initial-value representations in phase space employed in this work. These representations describe the system dynamics in phase space, and  are the Coupled Coherent States (CCS) method, which is a formally exact approach for describing quantum mechanics, and the Herman-Kluk (HK) propagator, which is semiclassical.
Both methods can be derived from the same source, and this derivation is provided in the Appendix. More details can be found in the review article \cite{Shalashilin_2004}.

The CCS method represents a time-dependent wave function as a superposition of time-dependent, nonorthogonal Gaussian coherent states (CS) $|z\rangle=|z(t)\rangle$ guided along the trajectory determined by the Hamiltonian averaged within such a basis.  Explicitly,
 \begin{equation}|\Psi (t)\rangle=\int |z\rangle D_{z}(t)\mathrm{e}^{\mathrm{i}S_z}\frac{\mathrm{d}^2z}{\pi},\label{eq:Psi_D}\end{equation}
where the coherent state $|z\rangle$ is labeled by a single complex number
\begin{eqnarray}
\eqalign{
z=\sqrt{\frac{\gamma}{2}}q+\frac{\mathrm{i}}{\sqrt{2\gamma}}p, \cr
z^{*}=\sqrt{\frac{\gamma}{2}}q-\frac{\mathrm{i}}{\sqrt{2\gamma}}p,
}\label{eq:CS}
\end{eqnarray}which is defined in terms of the position $q=q(t)$ and the momentum $p=p(t)$ of the particle.
The expression
\begin{equation}
S_{z}=\int\left[\frac{\mathrm{i}}{2}\left(z^{*}\frac{\mathrm{d}z}{\mathrm{d}t}-z\frac{\mathrm{d}z^{*}}{\mathrm{d}t}\right)-H_{\mathrm{ord}}(z^{*},z)\right]\mathrm{d}t\,\label{eq:Sz} ,
\end{equation}
denotes the classical action along the trajectory defined with regard to the matrix element
$H_{\mathrm{ord}}(z^{*},z)=\langle z|\hat{H}_{\mathrm{ord}}(\hat{a}^{\dagger},\hat{a})|z\rangle$. This  represents
the diagonal elements of the ordered Hamiltonian matrix $\hat{H}_{\mathrm{ord}}(\hat{a}^{\dagger},\hat{a})$. In general,
\begin{equation}
\langle z|\hat{H}|z^{\prime}\rangle=\langle z|z^{\prime}\rangle H_{\mathrm{ord}}(z^{*},z^{\prime}),\label{eq:Hord}
\end{equation}
where $|z\rangle$, $|z^{\prime}\rangle$ denotes two arbitrary coherent states.

In coordinate space,
\begin{equation}\langle x|z\rangle=\left(\frac{\gamma}{\pi}\right)^{1/4}\exp\left[-\frac{\gamma}{2}(x-p)^2+
\mathrm{i}p(x-q)+\frac{\mathrm{i}pq}{2}\right]\label{eq:ansatz}\end{equation}
is a Gaussian wave packet centered at the phase-space coordinates $q$ and $p$. %In this work, employ around $1600$ trajectories for the CCS method.

For the Herman Kluk propagator, the time-dependent wave function is once more expressed in terms of Gaussians in phase space 
(for seminal papers  see \cite{Herman_1984,Kluk_1986,Herman_1986,Kay_1994}). 
Following the original work \cite{Herman_1984}, they are  usually 
labeled not with a single complex number $z$ but with two real numbers $p$ and $q$. 
In the $p,q$-notation the phase of CS in (\ref{eq:ansatz_HK})
differs from that of $z$-notations by  $\mathrm{i}pq/2$ which is compensated by the different form of action 
(\ref{eq:action_HK}) used in \cite{Herman_1984} as opposed to that of Eq.~(\ref{eq:Sz}).
This specific formulation is considered in our implementation and in the results that follow.  Explicitly,
\begin{equation}
|\Psi_{\mathrm{HK}}(t)\rangle=\int\hspace{-0.2cm}\int%
|q,p\rangle R(t,q_0,p_0)\langle q_0,p_0|\Psi(0)\rangle\mathrm{e}^{\mathrm{i}
S_{\mathrm{cl}}}\frac{\mathrm{d} q_0\mathrm{d} p_0}{2\pi} \label{PsiHK}
\end{equation}
where $|q,p\rangle$ represents a coherent state whose expression in coordinate representation is given by
\begin{equation}
\langle x|q,p\rangle=\left(\frac{\gamma}{\pi}\right)^{1/4}\exp%
\left[-\frac{\gamma}{2}(x-q)^2+\mathrm{i}p(x-q)\right],\label{eq:ansatz_HK}
\end{equation}
which differs from Eq.~(\ref{eq:ansatz}) employed in the CCS approach by an insignificant phase factor, and
\begin{equation}
R(t,q_0,p_0)=\frac{1}{2^{1/2}}\left(m_{pp}+m_{qq}-\mathrm{i}\gamma m_{qp}+\frac{
\mathrm{i}}{\gamma}m_{pq}\right)^{1/2}\label{eq:prefactor_HK}
\end{equation}
is given in terms of the elements $m_{uv}=\partial u/\partial v_0$ of the monodromy matrix. In Eq.~(\ref{PsiHK}), the action reads as
\begin{equation}
S_{\mathrm{cl}}(q,p)=\int\left(p\dot{q}-H_{\mathrm{cl}}\right)dt, \label{eq:action_HK}
\end{equation}
where $H_{\mathrm{cl}}$ is the classical Hamiltonian, in which the operators $\hat{x}$, $\hat{p}$ in (\ref{eq:Hoperator}) have been replaced by the phase-space variables $q$, $p$.
For the initial wave packet (\ref{eq:initialwp}) considered here,
\begin{equation}
\hspace*{-0.8cm}\langle q_0,p_0|\Psi(0)\rangle=\exp\bigg\{-\frac{\gamma}{4}(q_{\alpha}-q_0)^2-\frac{1}{4\gamma}(p_{\alpha}-p_0)^2+\frac{\mathrm{i}}{2}(p_{\alpha}+p_0)(q_0-q_{\alpha})\bigg\}.
\label{eq:initialphasespace}
\end{equation}
The integral is carried out over phase space coordinates which are used as
initial conditions of the classical solutions $(q,p)$. 

Apart from trivial notations, the HK propagator and the CCS method differ in three main ways:
\begin{enumerate}

\item[(1)] The trajectories of the Gaussian Coherent States in the HK method are purely classical, while in 
CCS the trajectory of a CS $|z(t)\rangle$ is driven by the Hamiltonian $\langle z|\hat{H}|z\rangle$. 
This latter Hamiltonian is the average of the quantum Hamiltonian with regard to the coherent states, and takes into account the local zero-point energy and further corrections due to commutators. This makes all wells more shallow and lowers all potential barriers. The CCS and HK trajectories are identical only in the case of harmonic potential.

\item[(2)] The CCS representation is a formally exact basis set technique. It uses coupled quantum equations for 
the coefficients $D_z(t)$, obtained simply by substitution of (\ref{eq:Psi_D}) into the Schr\"odinger equation, 
to propagate the wave function (\ref{eq:Psi_D}). In contrast, in the HK theory the coefficients are obtained by an 
analytical semiclassical formula, which includes the elements of the monodromy or stability matrix. These expressions result from the so called local quadratic approximation, which only takes into account the first and second term in the Taylor expansion of the potential energy around a specific trajectory, and assume that only the coupling of coherent states near this trajectory is important. Physically, this implies that, while in the CS representation the trajectories are coupled through the amplitudes $D_z(t)$, each trajectory in the swarm employed in the HK method contributes independently. Indeed, for each trajectory there is one prefactor, which depends 
only on the information carried out by that specific trajectory.

\item[(3)] On a more technical level, the CCS is often used in conjunction with various algorithms of basis set expansion, 
which generate additional basis functions and reproject the wave function on the new basis set. 
These adoptive basis sets allow to follow complicated features of the dynamics more efficiently. 
Although reprojection has been used in conjunction with the HK method as well, in the latter case it is less common \cite{Shalashilin_Jackson_2000,Burant_Batista_2002}.
\end{enumerate}

In the Appendix a sketch is presented of how both CCS coupled equations and the HK formula are obtained from the same source, which is the integro differential form of the Schr\"odinger equation in the continuous CS representation, closely following the review article \cite{Shalashilin_2004}. For a detailed account of the similarities and differences between both initial-value representations, see also \cite{Grossmann_1998,Miller_2002,Child_2003}.
\section{Phase-space dynamics}

\label{phasespace}

In the following, we wish to analyze different ionization mechanisms in phase space. According to the quasi-static tunneling picture, a
low-frequency, time-dependent field and the binding potential determine an
effective potential barrier
\begin{equation}
V_{\mathrm{eff}}(x,t)=V_a(x)+x\mathcal{E}(t),  \label{eq:barriert}
\end{equation}
whose maximum will be given by $V_{\mathrm{eff}}(x_s,t)$ at the coordinate $%
x_s$ such that $\partial V_{\mathrm{eff}}(x,t)/\partial x|_{x=x_s}=0$. If
the total electron energy is larger than this value, it may escape via
over-the-barrier ionization. If the total energy is smaller, tunneling is
expected to be the dominant ionization mechanism.

\subsection{Phase portraits and classical-trajectory analysis}

With this aim in mind, we will perform a phase space analysis of both the trajectory
ensemble used to construct the semiclassical wave function in Eq.~(\ref%
{PsiHK}), and the wave functions $\Psi_{\mathrm{HK}}(x,t)$ and $\Psi(x,t)$. For
simplicity, we will first study what happens if the driving field
is static. This is a good approximation for the instantaneous barrier (\ref%
{eq:barriert}) if the frequency is low enough. In this case, the time-dependent field $\mathcal{E}(t)$ is replaced by
$\mathcal{E}$ in Eq.~(\ref{eq:laserpot}).

In phase space, the classical static Hamiltonian then reads as
\begin{equation}
H^{\mathrm{st}}_{\mathrm{cl}}(p,q)=\frac{p^2}{2}+V_a(q)+\mathcal{E}q.
\label{eq:Hstatic}
\end{equation}
In these studies, we will consider the soft-core potential (\ref{eq:softcore}).

\begin{figure*}[tbp]
\centering
\mbox{\subfigure{\includegraphics[width=2.5in]{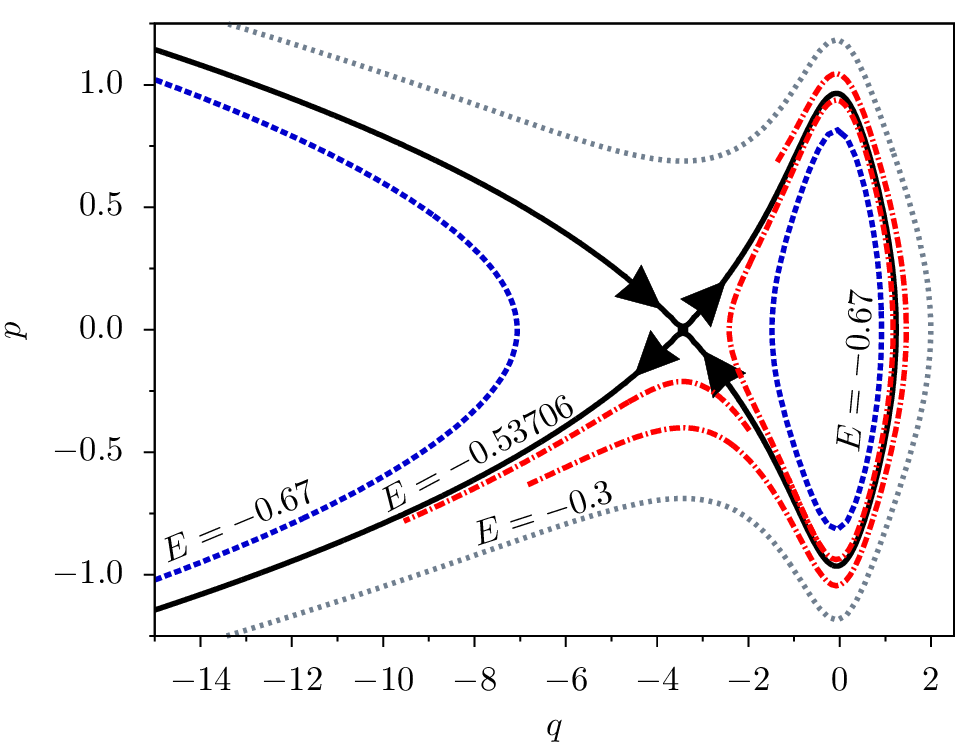}}\quad
\subfigure{\includegraphics[width=2.5in]{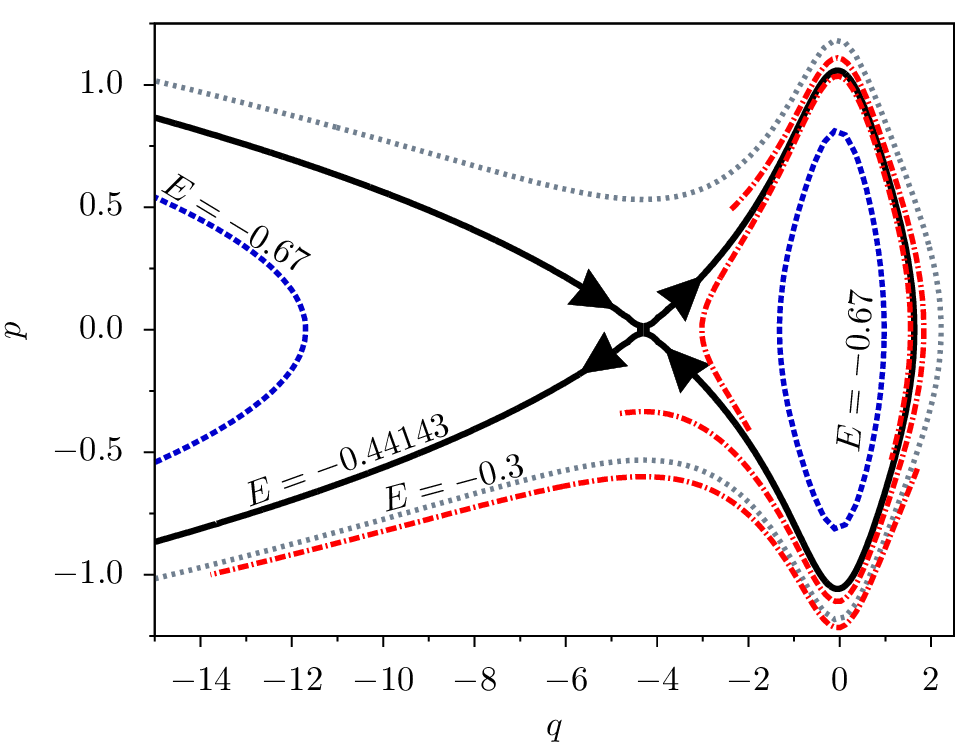}}}
\caption{Phase portrait of the system defined by~(\protect\ref{eq:Hstatic}).
Solid lines represent the separatrix in phase space for driving-field
amplitudes of $\mathcal{E}=0.075$ (left) and $\mathcal{E}=0.05$ (right).
Dashed and dotted lines show solutions for energies $E=-0.67$ and $E=-0.3$,
respectively. The dashed-dotted lines illustrate the evolution of some sample trajectories from $t=0$ to $t=20$ a.u.}
\label{fig:separatrix}
\end{figure*}

In Fig.~\ref{fig:separatrix}, we present the phase portrait of the system for the Hamiltonian (\ref{eq:Hstatic})
and static fields of different amplitudes. The figure shows that the
condition $p^2/2+V_a(q)+\mathcal{E}q=E_{\mathrm{min}}$, where $E_{\mathrm{min}%
} $ is the minimal energy necessary for the electron to undergo
over-the-barrier ionization defines a
separatrix, which crosses at $(q,p)=(q_s,0)$. For energies $E$ below the
separatrix energy the dynamics can be either bounded or unbounded depending
on whether the spatial coordinate of the orbit is larger or smaller than
that of the saddle point $(q_s,0)$. This means that an orbit with $E<E_{\mathrm{min}}$ will remain unbounded if its
initial spatial coordinate $q_0$ lies on the left-hand side of the saddle point, ant that it will remain
bound if $q_0$ lies on the
right-hand side of $q_s$ (Figure~\ref{fig:separatrix}).
On the other hand, orbits with $E>E_{\mathrm{min}}$ will remain unbounded regardless of the
initial value of their spatial coordinate. Furthermore, the
trajectories in the ensemble always respect the constraints dictated by
classical dynamics. This is explicitly shown by the thin lines in the figure, which illustrate
the time evolution of some sample trajectories. In fact, the trajectories whose energy lie
below $E_{\mathrm{min}}$ always remain bound and propagate along closed
orbits bounded by the separatrix. Those that follow the separatrix from below
($E_{\mathrm{min}}<E<0.3$ a.u.) go around the bound region in phase space, but
never cross this region.

\begin{figure*}[tbp]
\centering
\mbox{\subfigure{\includegraphics[width=2.5in]{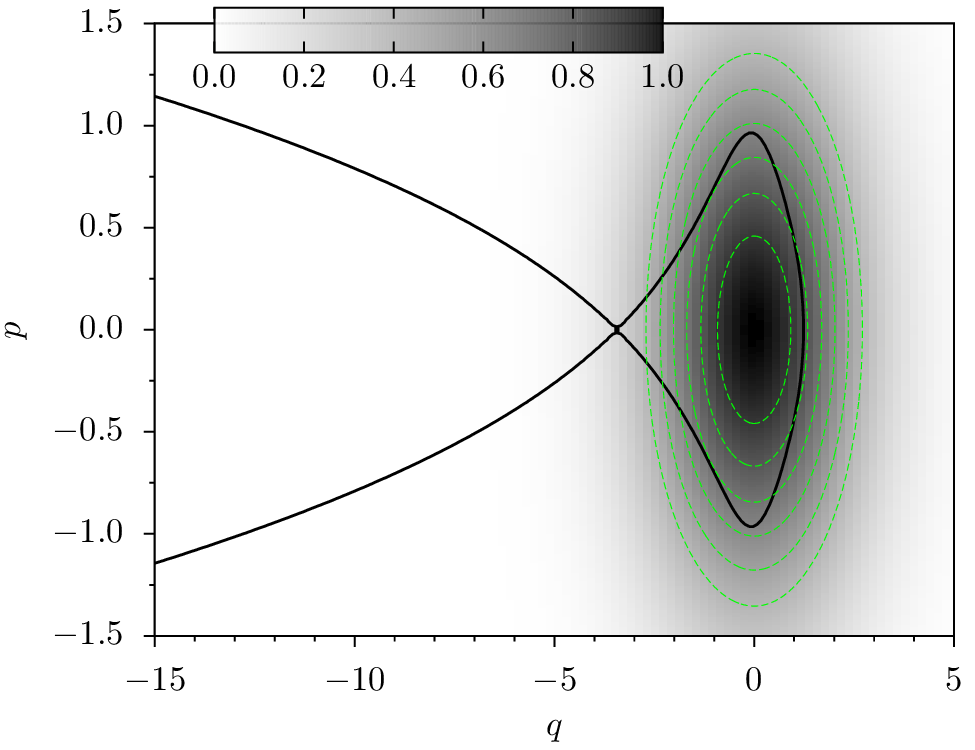}}\quad
\subfigure{\includegraphics[width=2.5in]{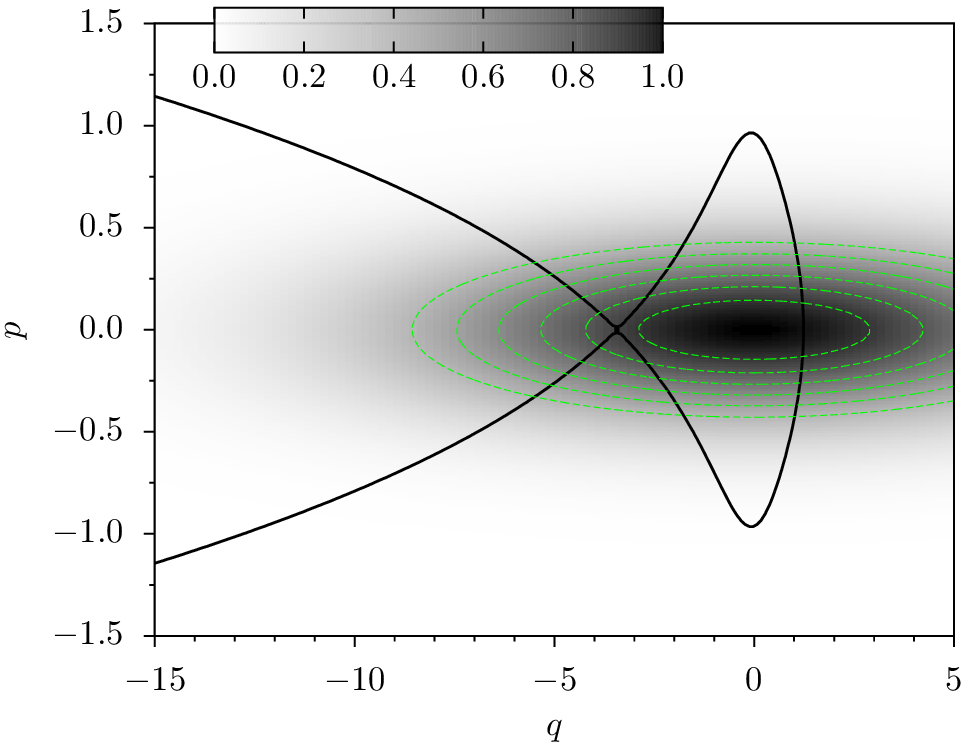}}}
\mbox{\subfigure{\includegraphics[width=2.5in]{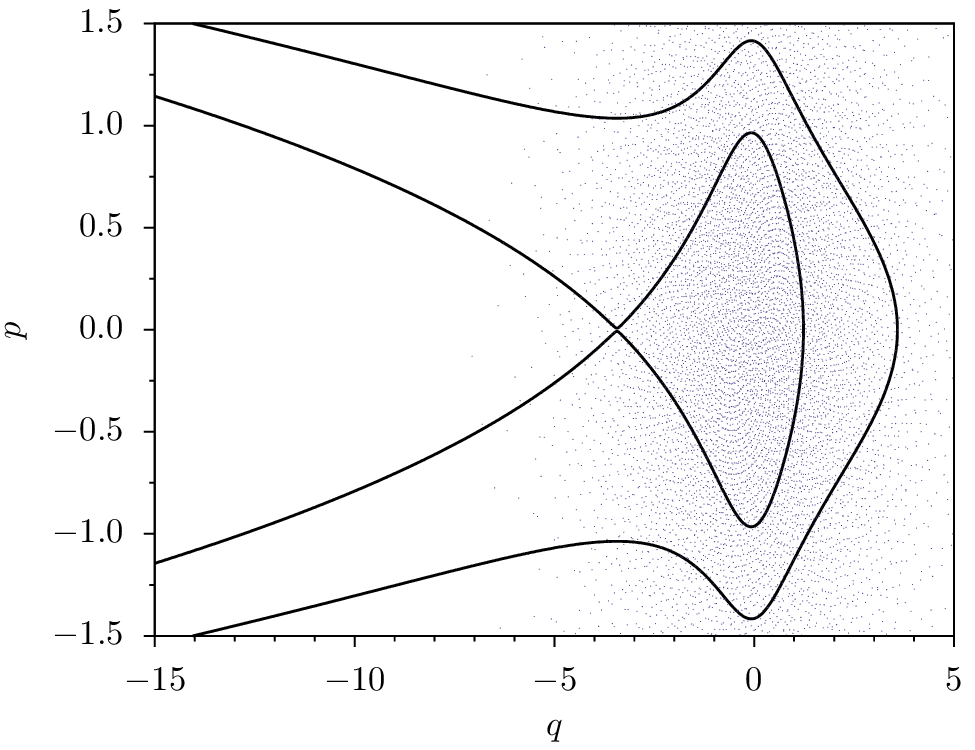}}\quad
\subfigure{\includegraphics[width=2.5in]{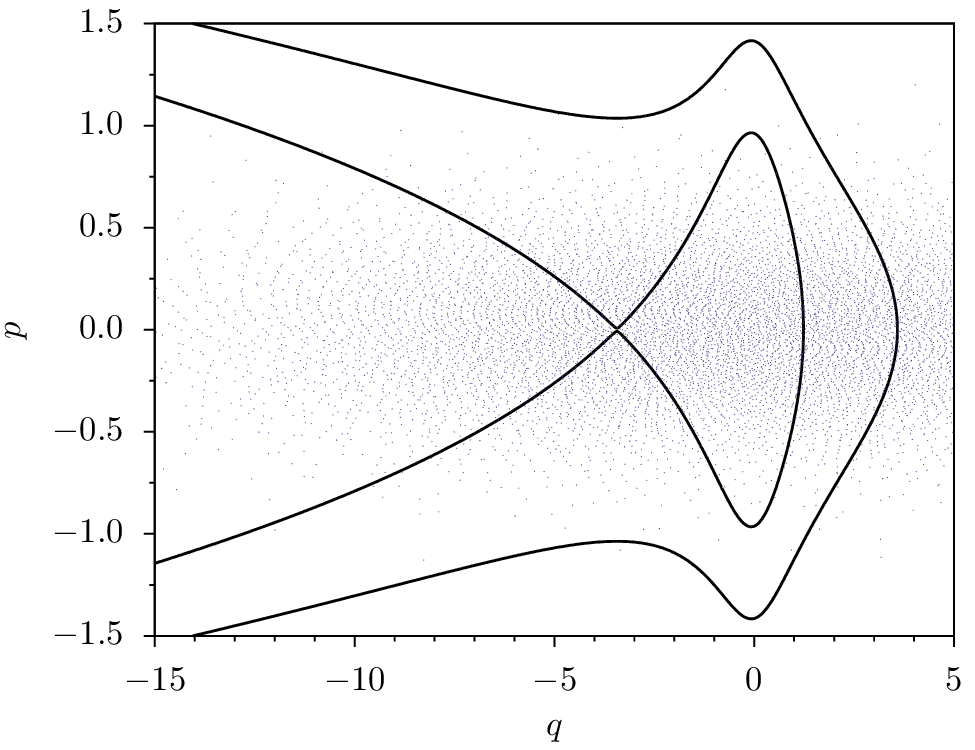}}}
\mbox{\subfigure{\includegraphics[width=2.5in]{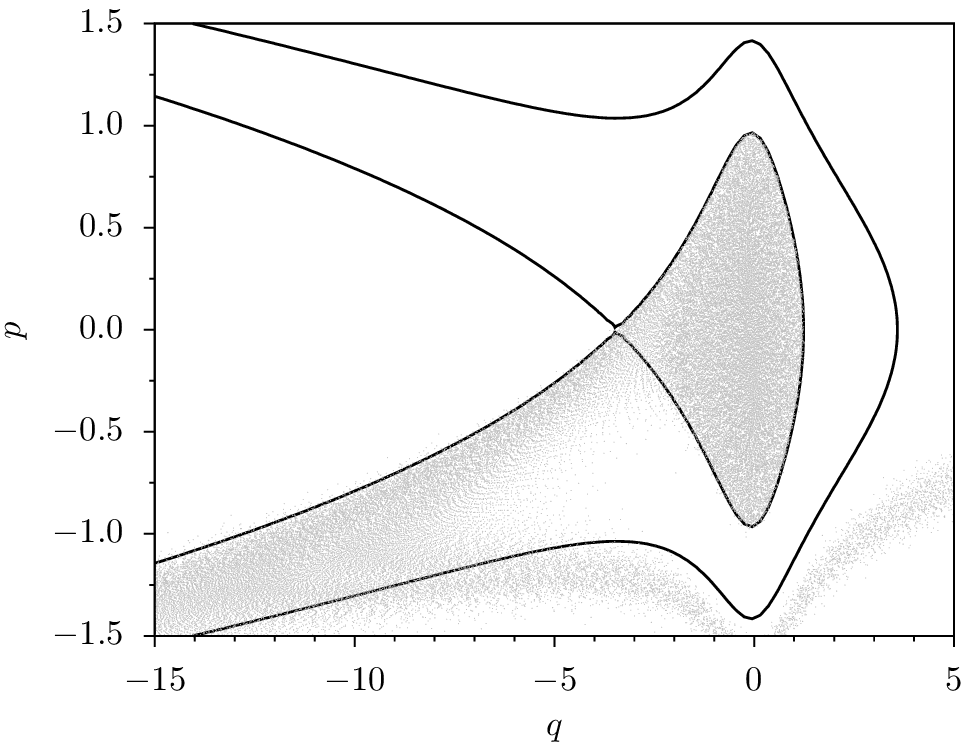}}\quad
\subfigure{\includegraphics[width=2.5in]{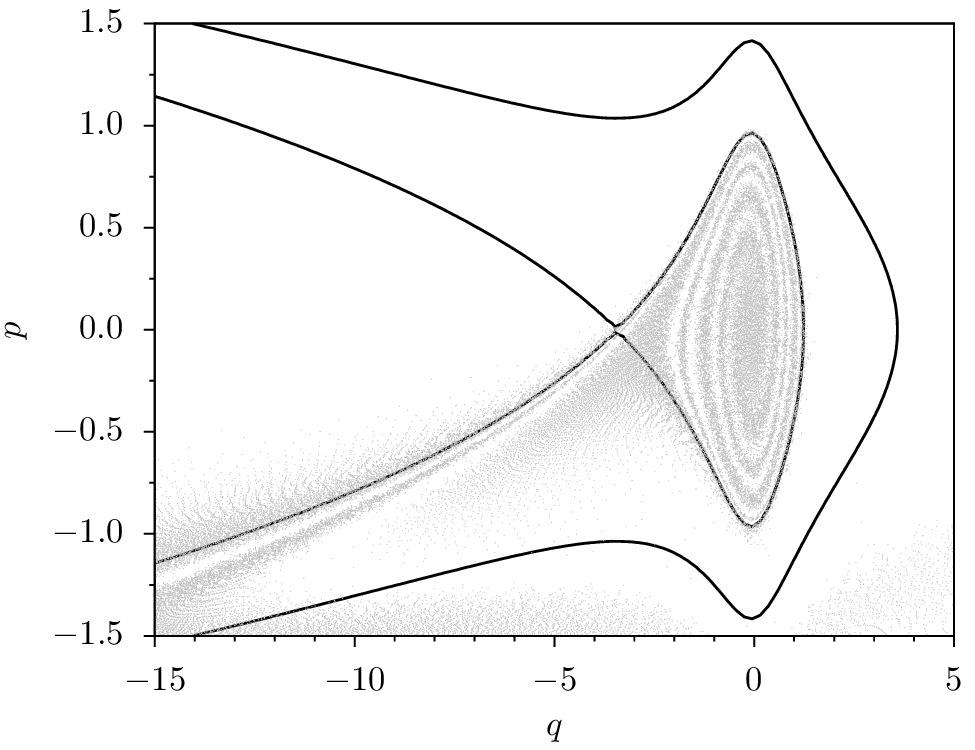}}}
\caption{Upper panels: Initial quantum mechanical distributions in phase
space for a Gaussian wavepacket centered at $(q_{\protect\alpha},p_{\protect%
\alpha})=(0,0)$ $\protect\gamma=0.5$ (bound-state energy $E\simeq-0.67$
a.u.; left) and $\protect\gamma=0.05$ (bound state energy $E\simeq-0.5$
a.u.; right). Middle panels: initial conditions in phase space for an
ensemble of classical trajectories corresponding to these distributions,
represented by dots. Lower panels: Classical distributions at $t=20$ atomic units for an ensemble
whose initial conditions are given by those in the middle panels [$(q_{\protect\alpha},p_{\protect%
\alpha})=(0,0)$ and, from left to right, $\protect\gamma=0.5$ = and $\protect\gamma=0.05$]. In our computations using the HK method, we have considered around $10^7$ trajectories.}
\label{fig:initialdist}
\end{figure*}

Classically, if a trajectory has a specific energy it will occupy a
well-defined phase-space orbit. Quantum mechanically, however, an initial
wave packet of a specific energy may occupy many regions in phase space. In
fact, due to the uncertainty relation, a strong spatial localization will
lead to a larger momentum spread and vice-versa. This is shown in the upper
panels of Fig.~\ref{fig:initialdist}, where we display the phase-space representation (\ref{eq:initialphasespace}) of two Gaussian wave
packets (\ref{eq:initialwp}) centered at $(q_{\alpha},p_{\alpha})=(0,0)$
with different widths $\gamma=0.5$ and $\gamma=0.05$. These widths give
bound-state energies of roughly $E=-0.5$ a.u. or $E=-0.67$ a.u.,
respectively. If now a set of initial conditions
in phase space is generated to match these distributions, these conditions will spread
over several regions, bound and unbound, as shown in the middle panels of
Fig.~\ref{fig:initialdist}. These are the starting points $(q_0,p_0))$ for an
ensemble of classical trajectories, which will be propagated in time and
used in the construction of the semiclassical wave function $\Psi_{\mathrm{HK}}(x,t)$%
.

After some time has
elapsed [lower panels in Fig.~\ref{fig:initialdist}], the distributions follow the constant-energy curves in phase space.
For the wave packet with $\gamma=0.5$, the distributions are concentrated in
the bound region or, for $q<q_s$, in the momentum region below the
separatrix. The region on the left-hand side of the saddle
around the axis $p=0$ is practically unpopulated. In contrast, for the wider
wave-packet in position space ($\gamma=0.05$), there are phase-space events
in this region, but closely following the separatrix from above.

These features are determined by the initial position and momentum spreads
of the wavepacket and the corresponding trajectory ensemble. For $\gamma=0.5$,
 the wavepacket is more localized in position space, so that the associated
classical ensemble practically does not occupy the region on the left-hand
side of the saddle. The trajectories that lie between the separatrix and the
curve associated with the energy $E=-0.3$ a.u. then follow the separatrix
from below. The remaining trajectories, for $E<E_{\mathrm{min}}$ or $E>-0.3$ a.u.
either remain bound or lead to the distribution below the latter curve, for
high negative momenta. In contrast, for a larger momentum spread ($%
\gamma=0.05$) the region on the left-hand side of the saddle is
reasonably populated from the start.

\begin{figure}[tbp]
\noindent
\begin{centering}
\includegraphics[angle=-0,scale=0.75]{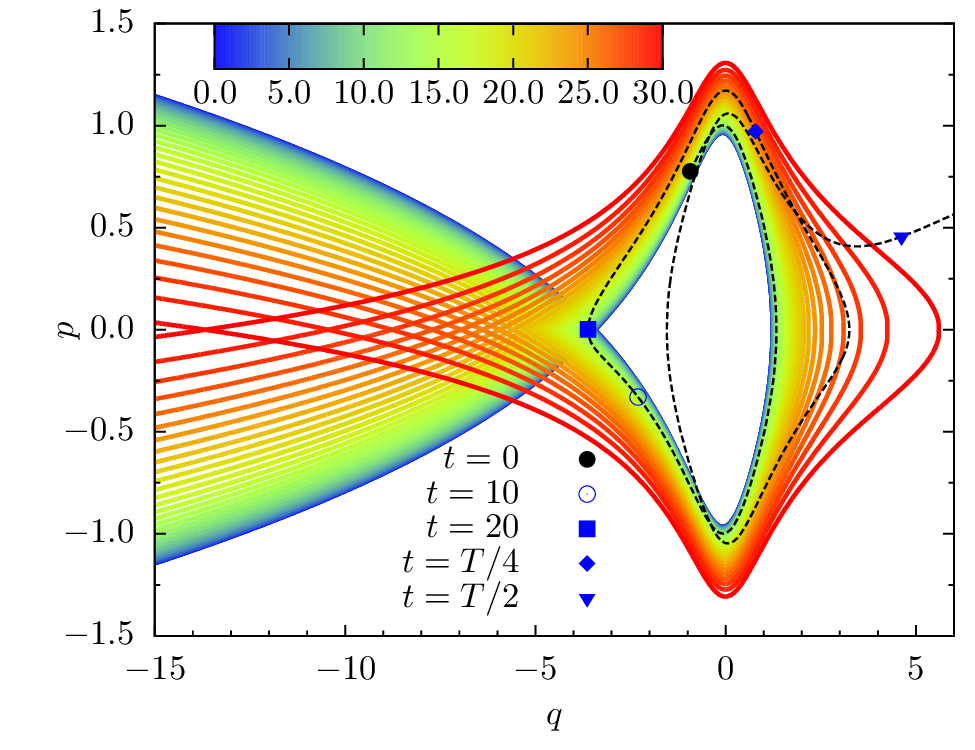}
\par\end{centering}
\noindent
\begin{raggedright}
\caption{\label{fig:phasesp3D} Time dependent separatrix corresponding to the laser field represented by Eq. (\ref{eq:TDPotential}) (solid lines), together with a specific electron trajectory (dashed line). The gradient color shows the time variation from $t=0$ (blue) to $t=30$ (red), which spans slightly less than the first quarter cycle of the the field. The dot depicts the initial condition of an initially unbound trajectory which becomes trapped because of the time dependent field, and the remaining symbols illustrate the phase-space coordinates $(p_t,q_t)$ at specific times $t$. For $t>T/4$, the region bounded by the separatrix starts to decrease, until the trapped trajectory is eventually able to escape.}
\par\end{raggedright}
\end{figure}

According to this analysis, only if the above-stated constraints vary in time may a classical bound trajectory become
unbound or vice-versa. This is what happens if the external field is time dependent.
In Fig.~\ref{fig:phasesp3D}, we illustrate this behavior for the laser
field associated with Eq.~(\ref{eq:laserpot}). In this case, the bound region first
increases in time, so that an initially unbound trajectory may become
trapped. After a field crossing, the field amplitude starts to increase,
and, consequently, the bound region will shrink. This will lead to a bound
trajectory becoming unbound. Hence, the electron will be able to escape.
According to the classical constraints, however, its momentum must be such
that it only may go over the barrier.

\subsection{Wigner quasiprobabilities}

\begin{figure*}[!htb]
%\noindent
\begin{centering}
\mbox{\subfigure{\includegraphics[width=3in]{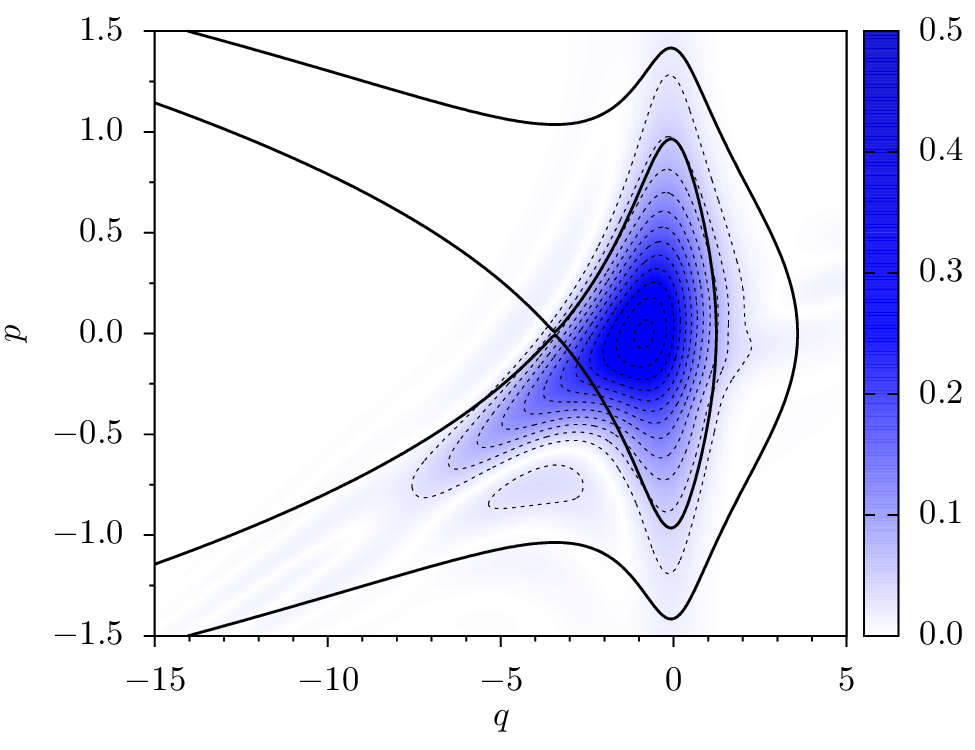}}\quad
\subfigure{\includegraphics[width=3in]{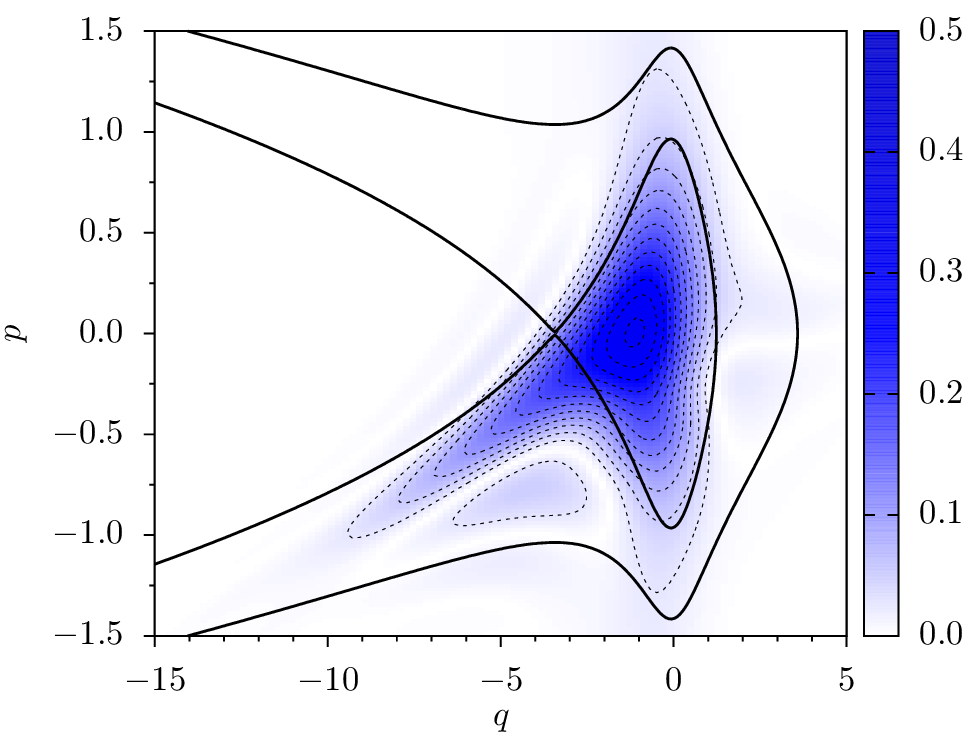}}}
\mbox{\subfigure{\includegraphics[width=3in]{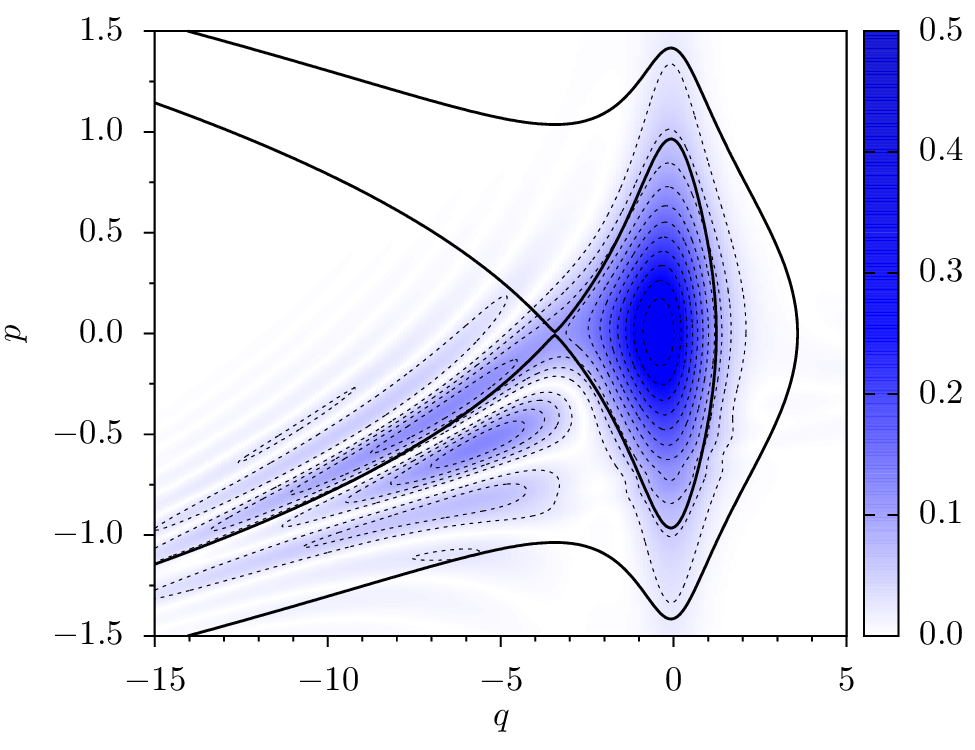}}\quad
\subfigure{\includegraphics[width=3in]{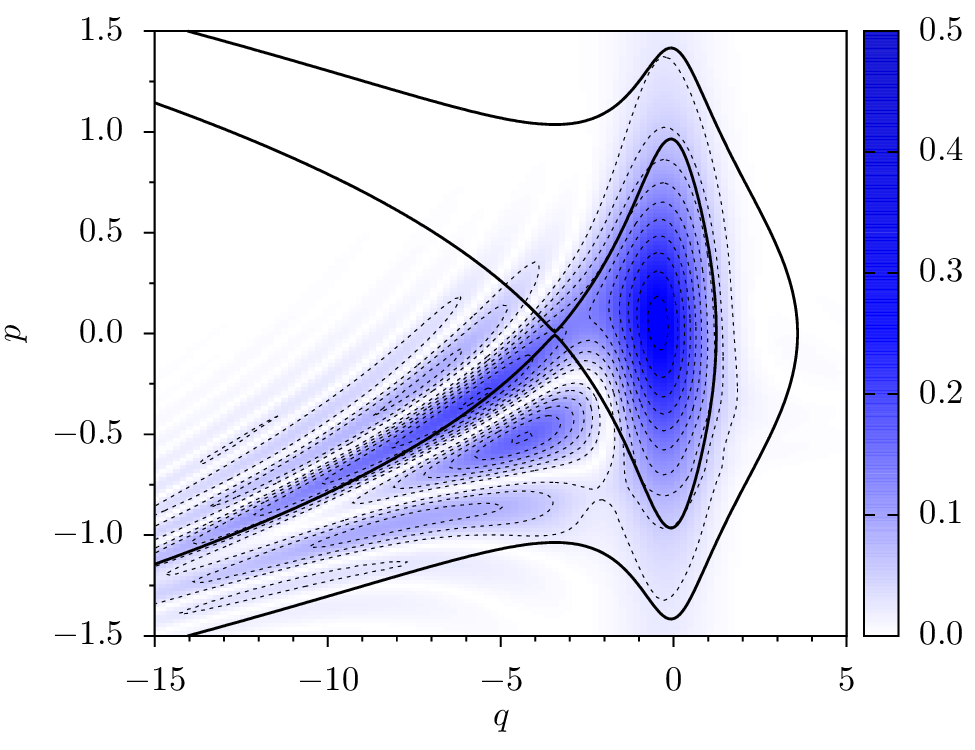}}}
\par\end{centering}
\noindent
\begin{raggedright}
\caption{\label{fig:wf1} Square of the Wigner function calculated for a wave packet in a static field of amplitude ${\cal E}=0.075$ a.u. by using the full quantum wave function (left column) and the semiclassical wave function (right column) for $t=10$ a.u. (first row) and $t=20$ a.u. (second row). Thicker lines show the separatrix and the phase space trajectory for $E=0$. The width and initial momentum of the initial wave packet are $\gamma=0.5$ and $p_{\alpha}=0$, respectively.}
\par\end{raggedright}
\end{figure*}
We will now employ the Wigner quasiprobability distribution (also known as
the Wigner function) in order to relate the above picture to the wave-packet
propagation in phase space \cite{Wigner_1932}. This function is defined as
\begin{equation}
W(q,p)=\frac{1}{\pi}\int^{+\infty}_{-\infty}\mathrm{d}y \Psi^*(q+y) \Psi(q-y) \exp[2\mathrm{i}py]  \label{Wigner}
\end{equation}
If Eq.~(\ref{Wigner}) is integrated over the momentum or position space, the corresponding
probability densities are recovered. One should note, however, that
the Wigner distribution may exhibit positive as well as negative values.
Hence, strictly speaking, it cannot be associated with a probability
density. Nonetheless, it does give an intuitive picture of the wave-packet
dynamics in phase space, and provide valuable information about momentum-position correlation. Wigner distributions are widely employed in quantum
optics, and have also been used in strong-field physics in order to access
the dynamics of ionization \cite{Czirjak_2000}, rescattering \cite%
{Kull_2012,Czirjak_2013}, double ionization \cite{Lein_2001} and HHG \cite%
{Graefe_2012}. A common feature identified in \cite{Czirjak_2000} and \cite{Graefe_2012} was a tail in the Wigner distribution, which has been associated with tunnel ionization. In the following, we will plot the square of the Wigner function, $W^2(q,p)$ as it makes the above-mentioned tail slightly clearer.

In the left and right panels of Fig.~\ref{fig:wf1}, we depict the squares of the Wigner
distributions obtained using the TDSE and the HK propagator, respectively,
for a static field and the soft-core potential (\ref{eq:softcore}).
This quasiprobability has evolved from an initial
Gaussian state centered at $(q_{\alpha},p_{\alpha})=(0,0)$ and width $%
\gamma=0.5$ (see Fig.~\ref{fig:initialdist} for details). The time
propagation of the wave packet, and in particular the shape of the Wigner function, are strongly influenced by the separatrix.
Throughout, the figure shows a distinct tail in the Wigner functions leaving the bound phase-space region. For short times, this tail
follows the separatrix from below, as shown in the upper panels of the
figure. This strongly suggests that the continuum is reached by
over-the-barrier ionization: the electronic wave packet does not leave the
core with vanishing momentum, but, rather, with the minimum necessary
momentum to overcome $V_{\mathrm{eff}}(q_s)$. As time progresses,
interference fringes start to build up on the left-hand side just above the
separatrix. These fringes have been identified in \cite{Czirjak_2000}, for a
delta-potential model in a static field, and in \cite{Kull_2012,Graefe_2012}
for long- and short-range potentials in time-dependent fields, and have been
associated with the quantum interference of ionization processes occurring
at different times. Apart from that, there is a pronounced tail now
following the separatrix from above. This implies that parts of the wave
packet are reaching the continuum with lower energy than that required for
over-the-barrier ionization to occur. In other words, tunnel ionization may
be taking place, both for the TDSE and the HK wave packets. Near $q=q_s$, the
momentum at this tail is even approximately vanishing, in agreement with
the model in \cite{Czirjak_2000}. A decrease in the momentum associated with the tail is intuitively expected as, physically, the components of the wave packet with lower energies will take longer to reach the barrier \cite{Maitra_1997}.

\begin{figure}[!htb]
\noindent
\mbox{\subfigure{\includegraphics[width=2.5in]{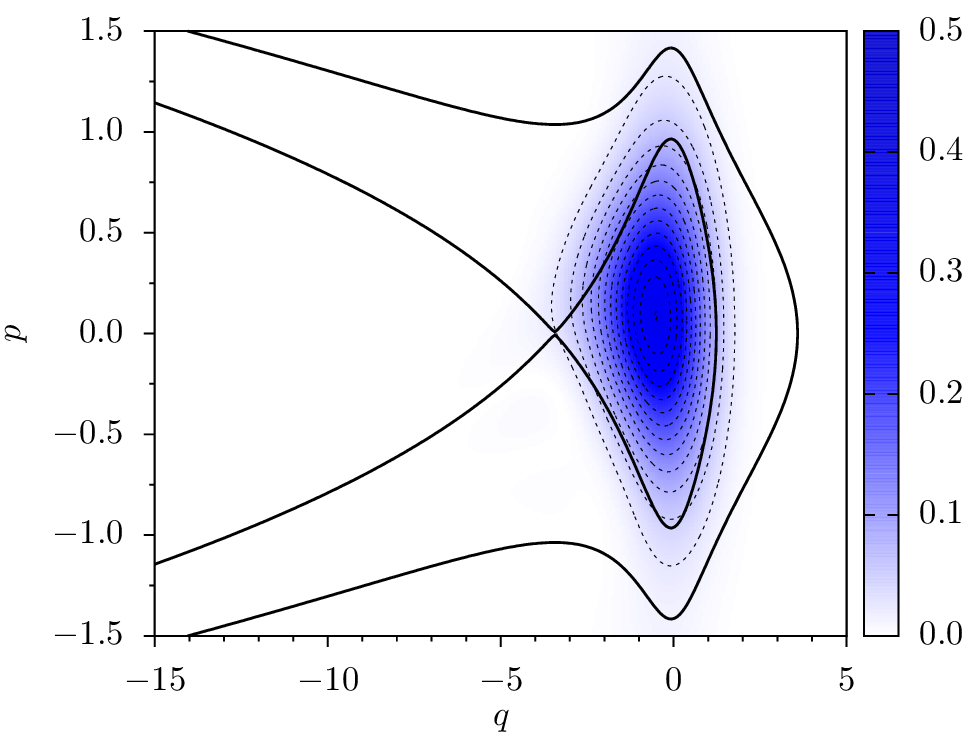}}\quad
\subfigure{\includegraphics[width=2.5in]{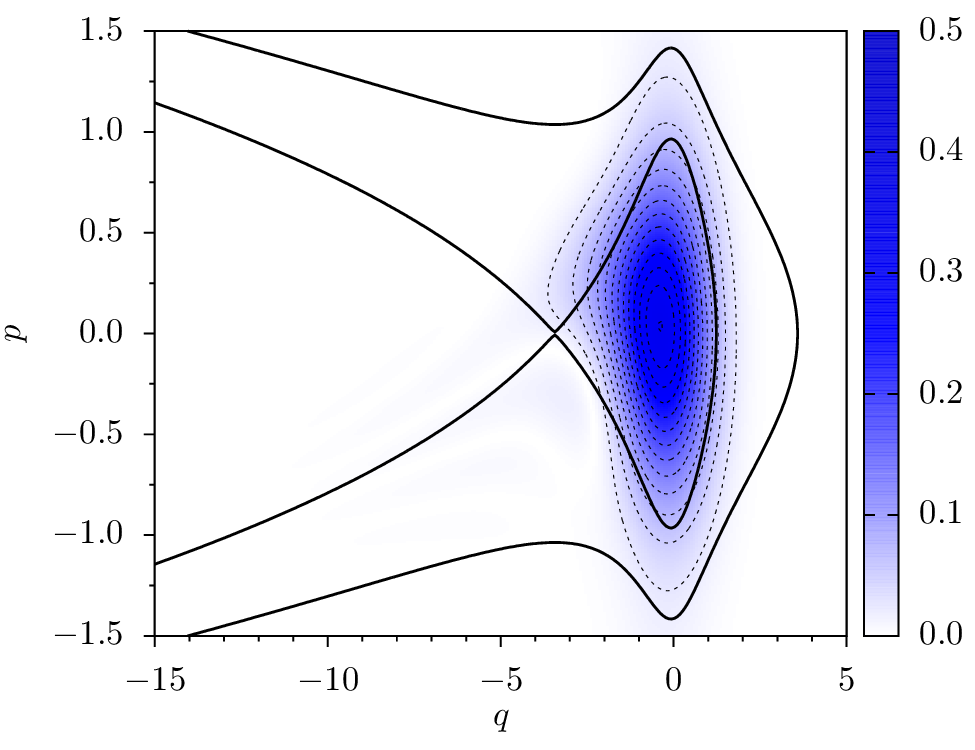}}}
\caption{\label{fig:wf_noionization} Square of the Wigner function calculated
using the HK propagator leaving out the trajectories starting outside the bound region, for the same parameters
as in Fig.~\ref{fig:wf1} and times $t=10$ a.u. and $t=20$ a.u. (left and right panels, respectively).}
\end{figure}
All the above-mentioned features are present both for the semiclassical and
quantum mechanical computations. In fact, the agreement between the outcomes
of both approaches is quite striking. The presence of interference patterns
in both cases is not surprising, as it is well known that the HK propagator
is capable of reproducing such effects.
It seems, however, that the semiclassical propagator allows the wavepacket to cross classically forbidden regions. In other words, classical trajectories corresponding to over-the-barrier energy appear to mimic transmission through a barrier for $E<E_{\mathrm{min}}$. This is not obvious, as these trajectories do not
violate any phase-space constraints, or cross classically forbidden regions.

\begin{figure}[!htb]
\begin{centering}
\mbox{\subfigure{\includegraphics[width=2.5in]{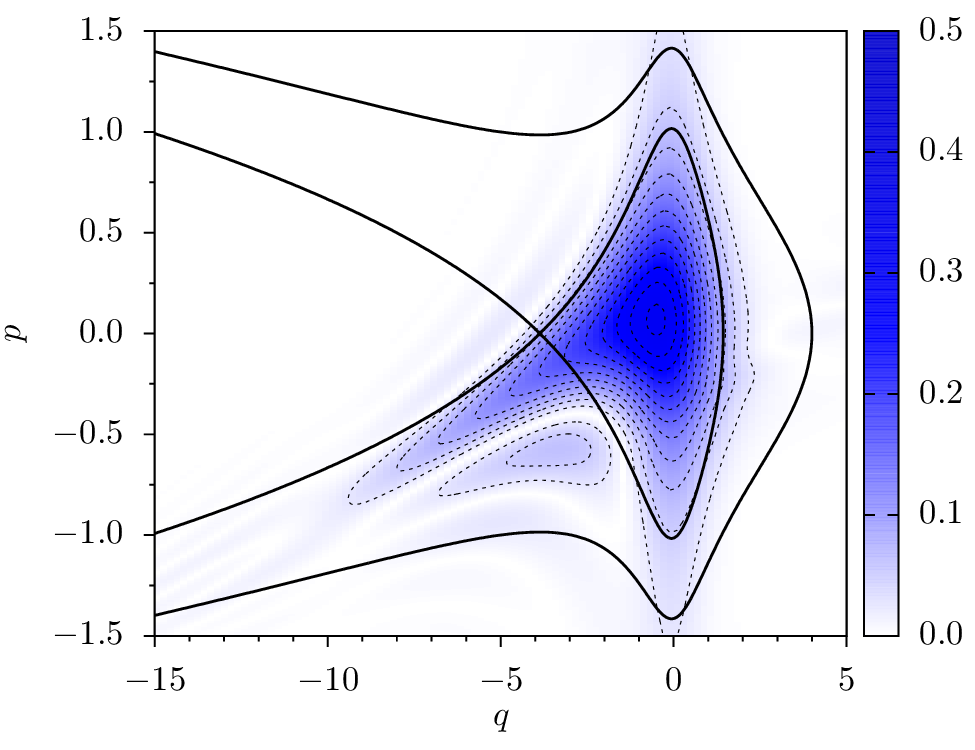}}\quad
\subfigure{\includegraphics[width=2.5in]{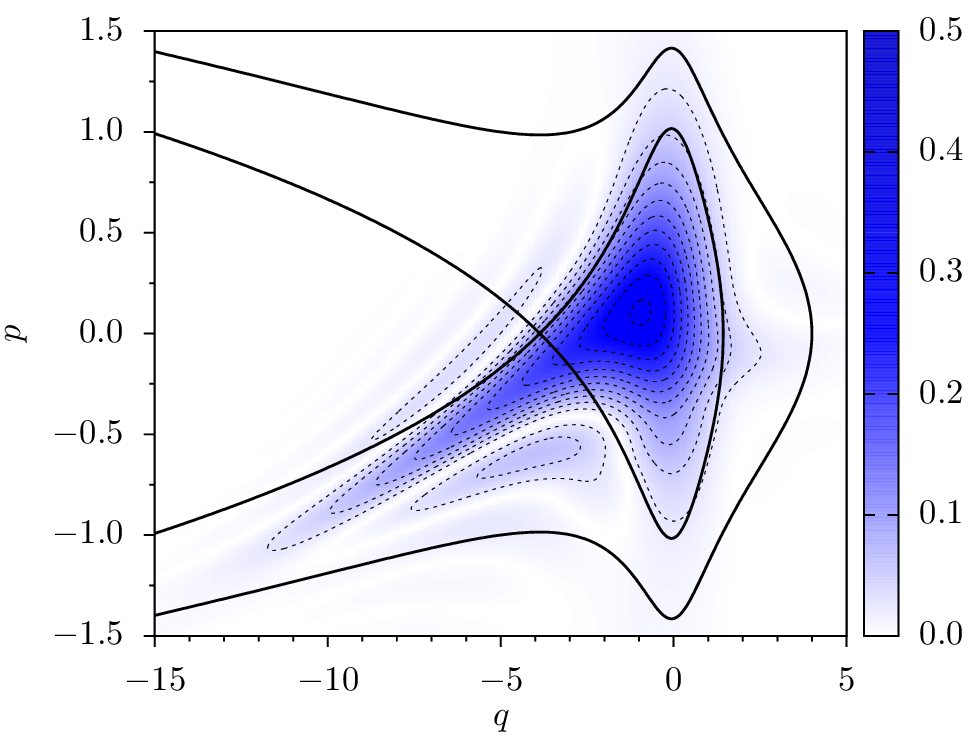}}}
\mbox{\subfigure{\includegraphics[width=2.5in]{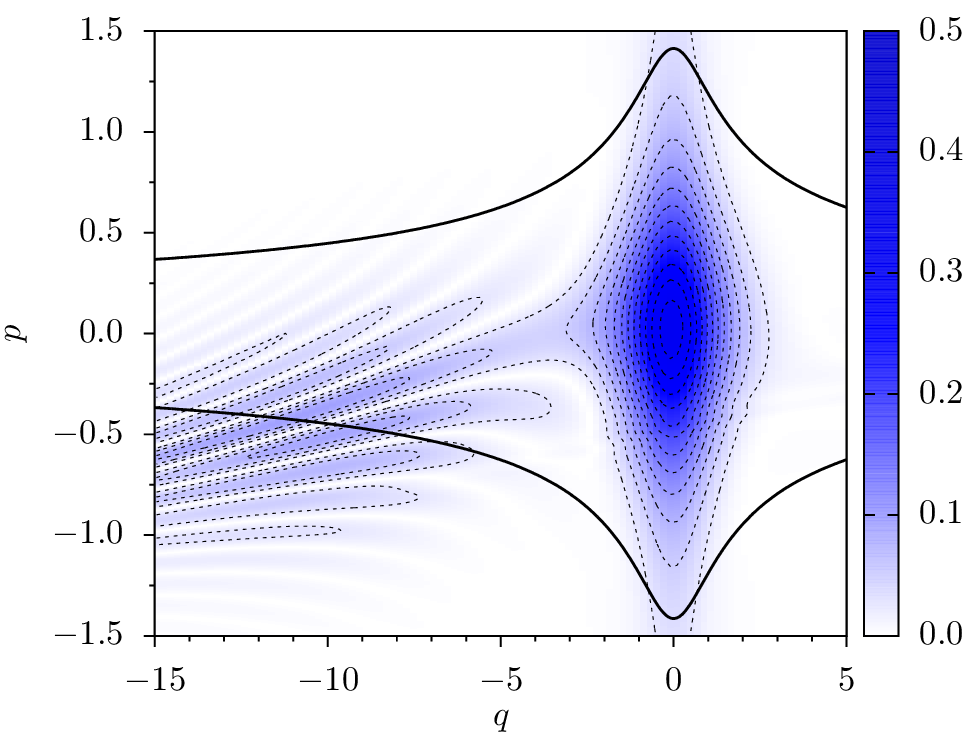}}\quad
\subfigure{\includegraphics[width=2.5in]{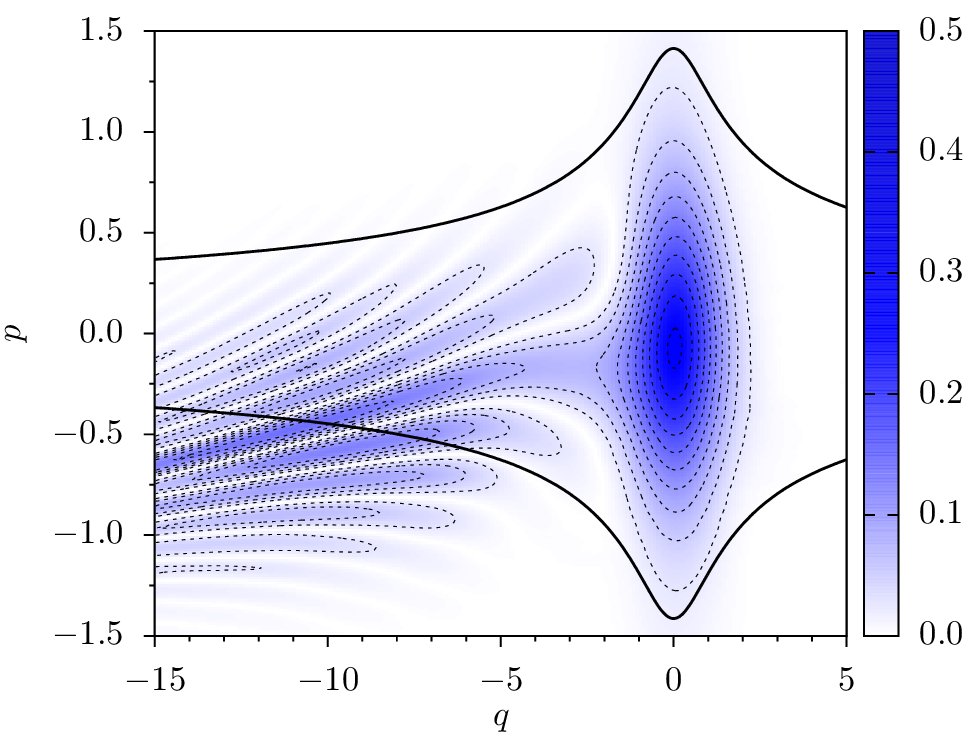}}}
\mbox{\subfigure{\includegraphics[width=2.5in]{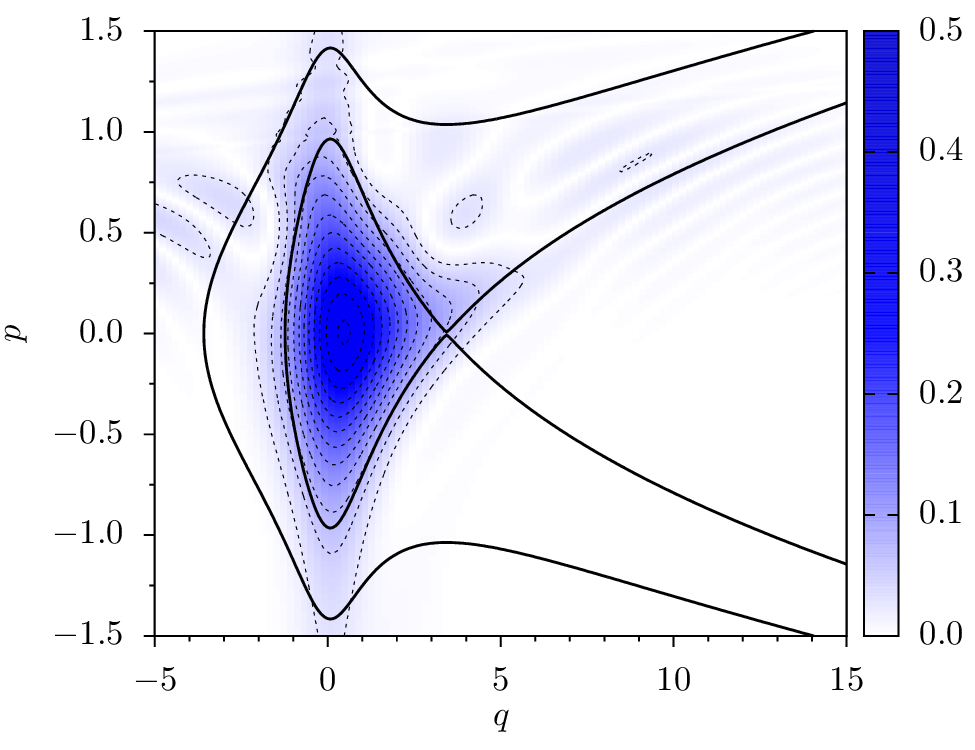}}\quad
\subfigure{\includegraphics[width=2.5in]{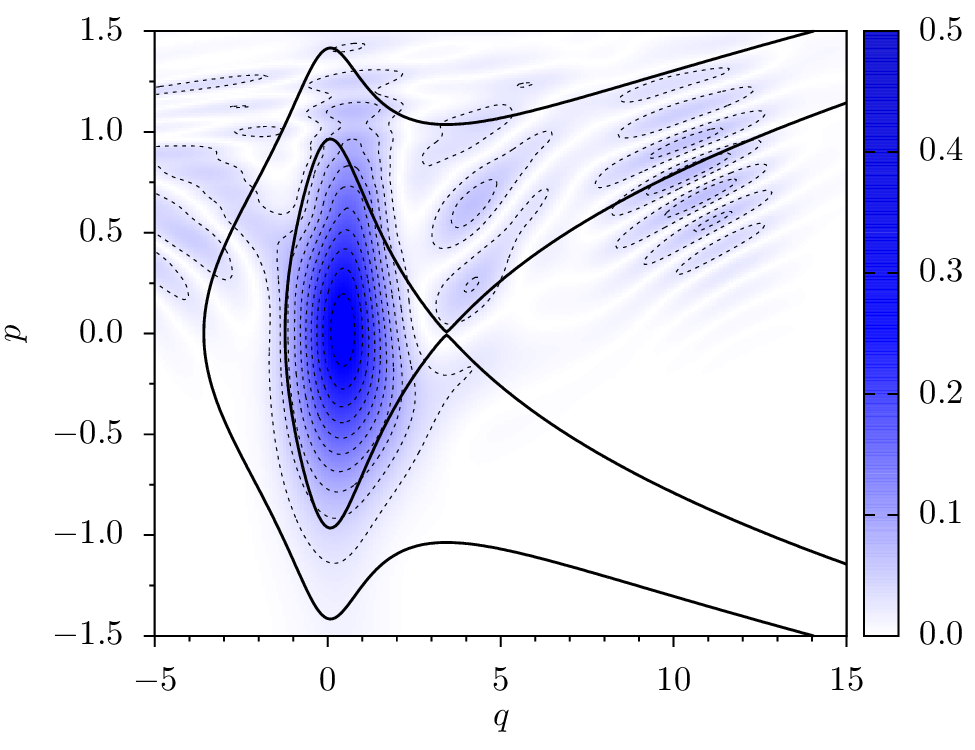}}}
\par\end{centering}
\noindent
\caption{\label{fig:wf_tdepend} Square of the Wigner function computed using the TDSE and the HK propagator (left and right panels, respectively) for a time-dependent field associated to  Eq.~\ref{eq:laserpot}, with frequency $\omega_0=0.05$ a.u. and amplitude $\mathcal{E}=0.075$ a.u. The top, middle and bottom panels have been calculated for  $t=12.6$, $t=T/4$ and $t=T/2$, respectively, where $T=2\pi/\omega_0$ is the field cycle.}
\end{figure}
We have however verified that the trajectories
whose initial coordinates $(q_0,p_0)$ lie outside the bound phase-space region lead to the tail
in the Wigner distributions. For clarity, in Fig.~\ref{fig:wf_noionization} we show
the Wigner function computed from the HK propagator leaving out these trajectories. In the figure, both the above-mentioned tail and the interference fringes are absent. These findings are in agreement with the results in \cite{Balazs_1990}, in which a nonlocal behavior of the Wigner function around separatrices has been observed for an inverted harmonic oscillator, and with those in \cite{Heim_2013}, which find that the transmission coefficient associated with a parabolic barrier is related to the quantum-mechanical weight of all classical trajectories with enough energy to go over the barrier. The disappearance of the fringes in the Wigner functions around $p=0$ for $q<q_s$ also support the assumption that the Wigner function is non-local.

For the time-dependent field in Eq.~(\ref{eq:laserpot}), we observe that the
tails with momenta following the separatrix from below prevail in the Wigner function. Furthermore, there are many
more fringes associated with quantum interference and rescattering events
(for discussions of these fringes see \cite{Kull_2012,Czirjak_2013}).
Snapshots of $W^2(q,p)$ in the time dependent case are presented in
Fig.~\ref{fig:wf_tdepend}.
The separatrix, which is now time dependent, is also displayed in the figure as the thick black lines. This suggests that, for this type of potential, there will be enough over-the-barrier dynamics for the approximate modeling of strong-field ionization and rescattering. We see, however, that the agreement between the HK propagator computation and the TDSE worsens with time. This is related to the fact that non-classical effects such as tunnel ionization and over-the-barrier reflections cannot be fully accounted for, and become dominant for longer times \cite{Maitra_1997}.
\subsection{Comparison with the CCS method}
\label{comparison}
We will now investigate the Gaussian potential (\ref{eq:gaussian}), and an initial wave packet
with $\gamma=1$.
For a short-range potential, the effective potential barrier is much
steeper and there is no Rydberg series. Hence, tunneling is expected to be
dominant for the parameter range in question. Furthermore, we will also perform a direct comparison with the CCS method,
which is well known to incorporate tunneling.

\begin{figure*}[tbp]
\centering
\mbox{\subfigure{\includegraphics[width=5cm]{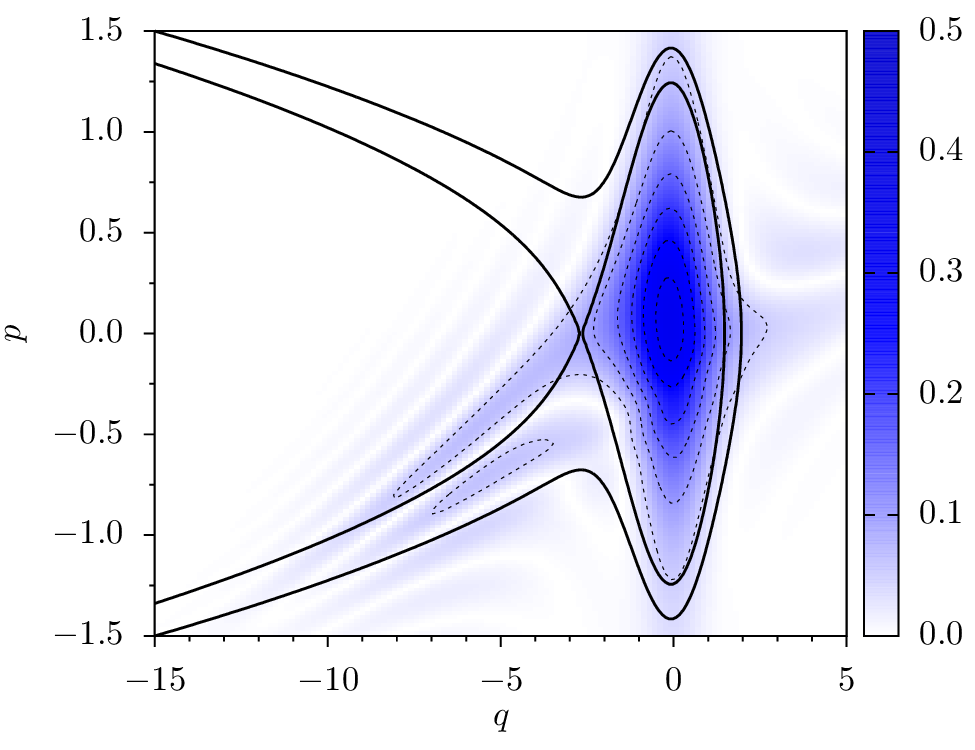}}\quad
\subfigure{\includegraphics[width=5cm]{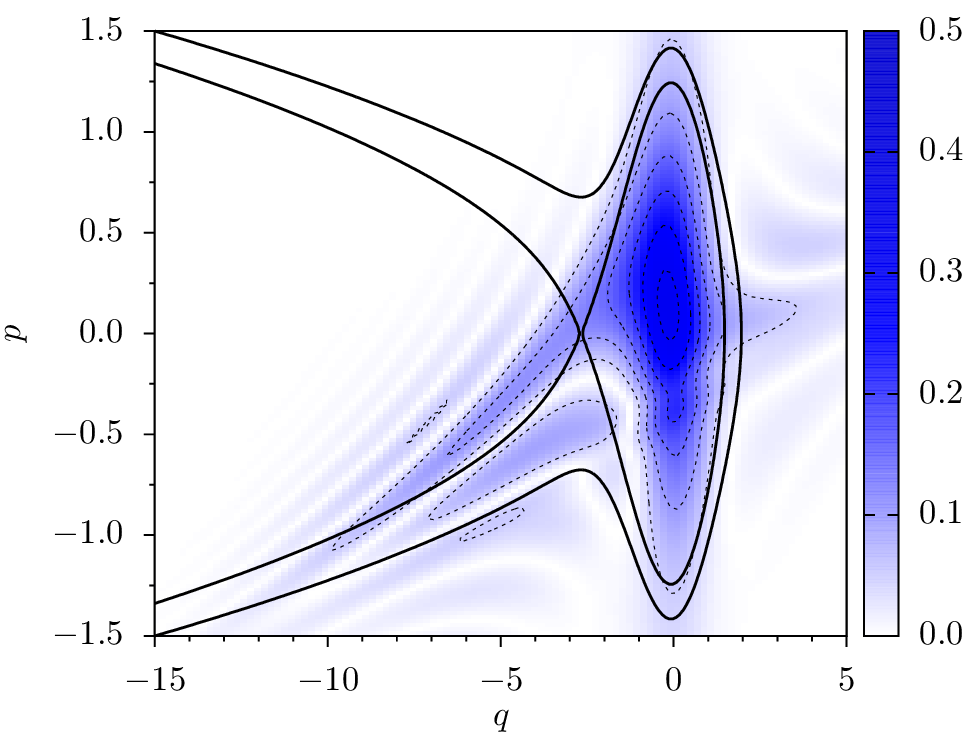}}\quad\subfigure{\includegraphics[width=5cm]{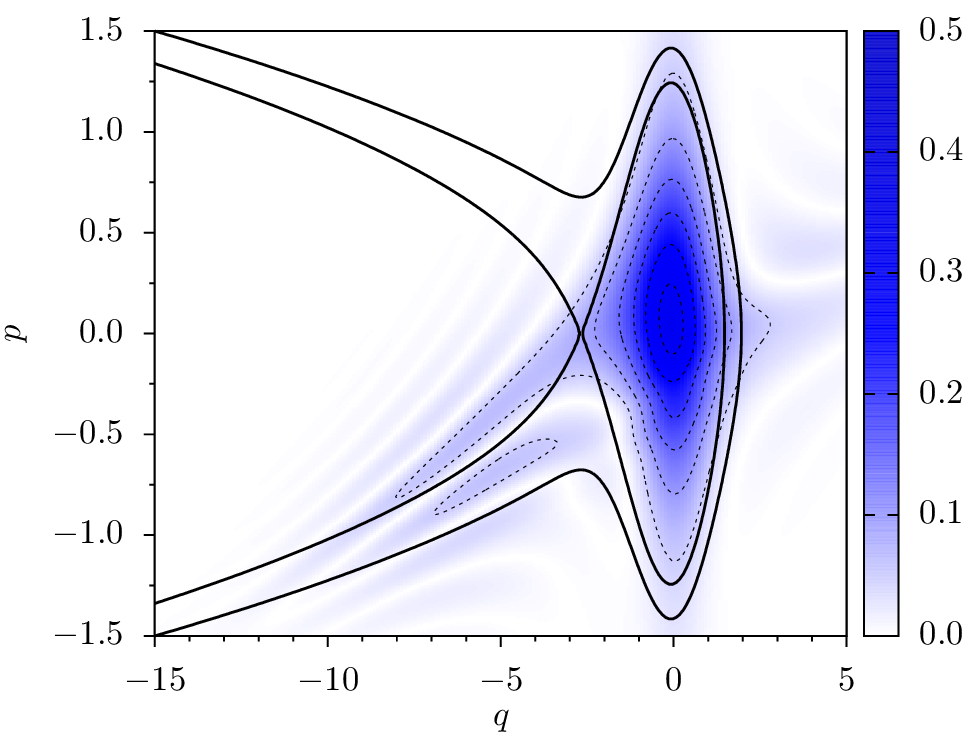}}}
\mbox{\subfigure{\includegraphics[width=5cm]{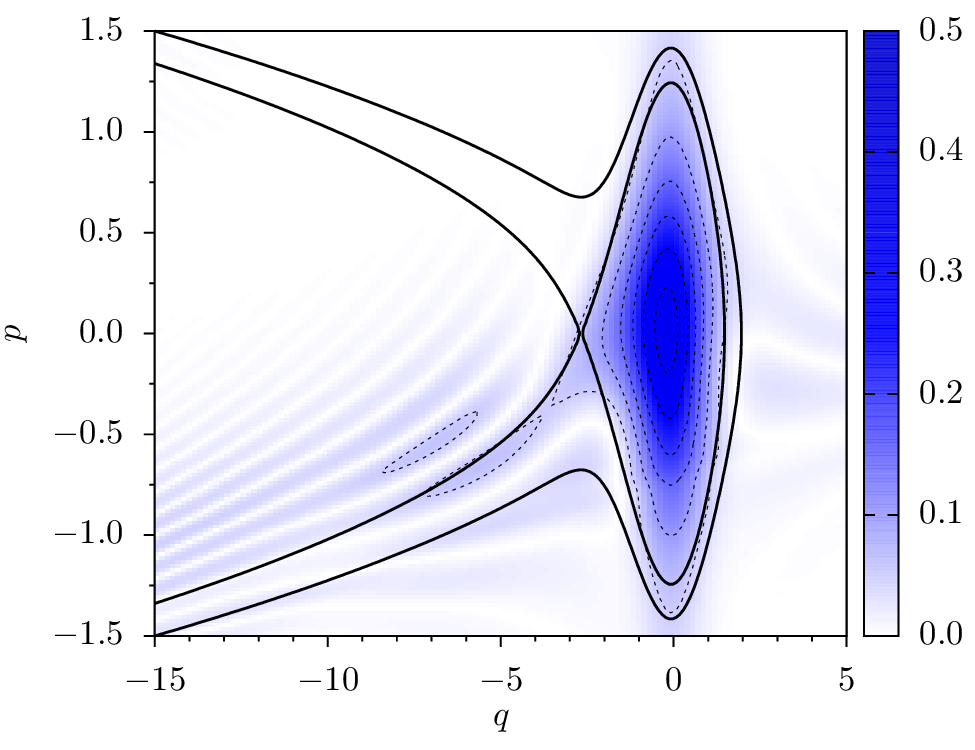}}\quad
\subfigure{\includegraphics[width=5cm]{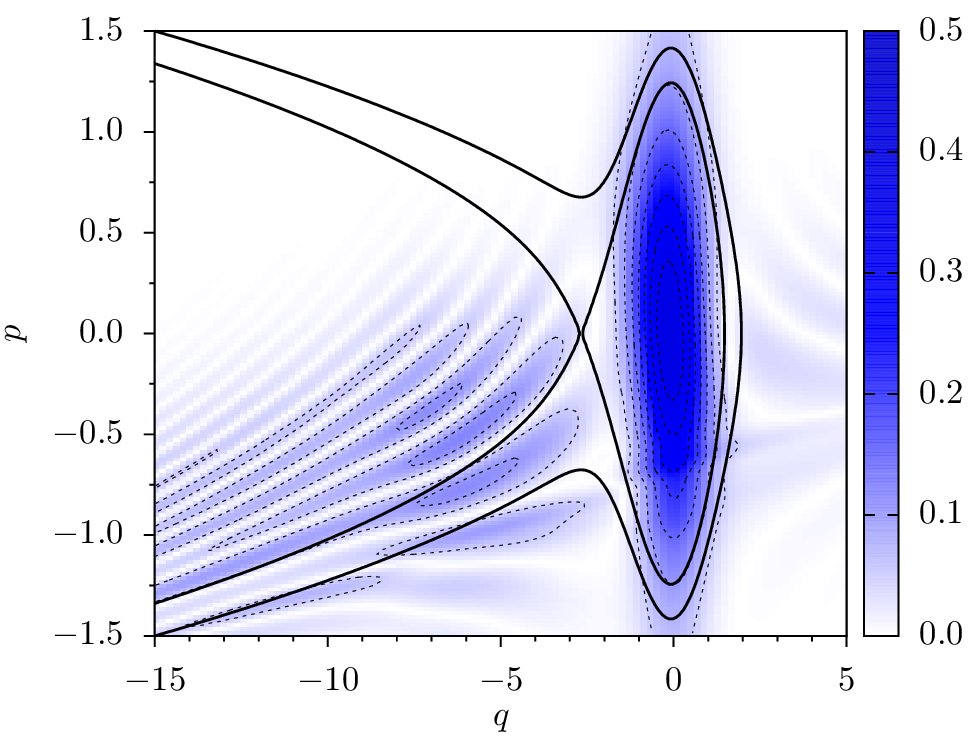}}\quad\subfigure{\includegraphics[width=5cm]{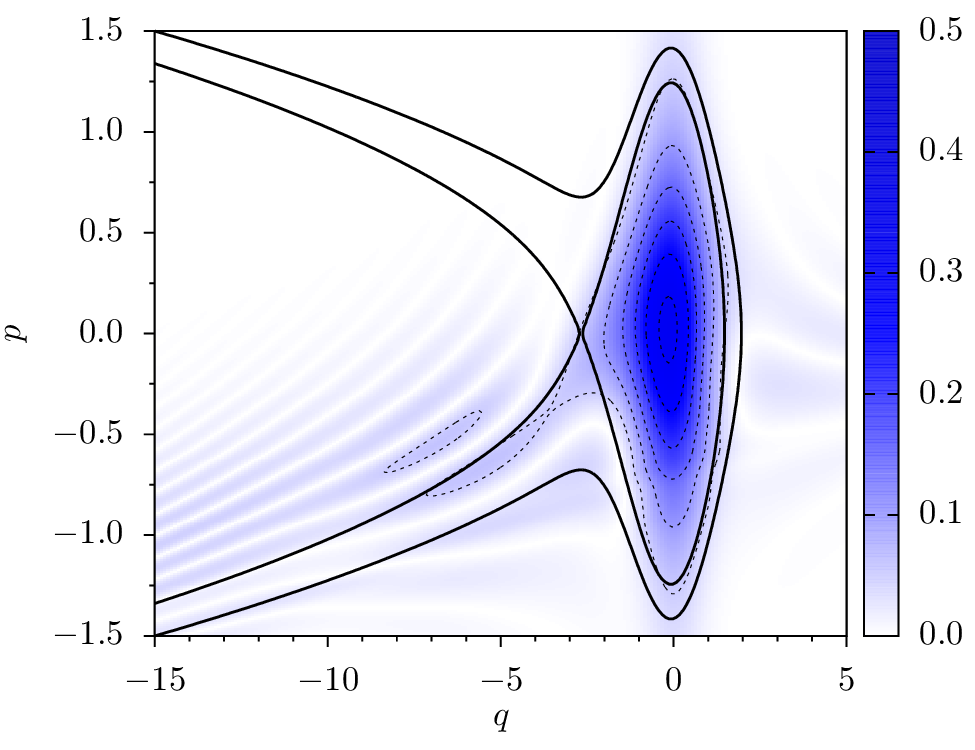}}}
\caption{Modulus square of the Wigner quasiprobability distributions computed using a static field of the same amplitude as in Figs.~\ref{fig:wf1} and the Gaussian potential (\ref{eq:gaussian}), using the TDSE, the HK propagator and the CCS method (left, middle and right panels, respectively). The upper and lower panels correspond to $t=10$ a.u. and $t=20$ a.u., respectively. The separatrix and the curve in phase space for the energy $E=0$ are illustrated by the thick lines in the figure. For the HK propagator and the CCS method we employ $10^{7}$ and
$1600$ trajectories, respectively . }
\label{fig:wignergaussian}
\end{figure*}

In Fig.~\ref{fig:wignergaussian}, we display snapshots of $W^2(q,p)$ computed with the TDSE, the HK propagator, and the CCS method
(left, middle and right columns, respectively) using a static field. For $t=10$
a.u. (upper panels) there is a very good
agreement between the outcome of all approaches, with a distinct tail in the quasiprobability
distributions following the separatrix from above. This behavior holds even for the results obtained with the
semiclassical propagator, and is different from that observed for the soft-core potential (see Fig.~\ref{fig:wf1}).
Physically, this suggests that there will be a nonvanishing probability density leaving the core with
$|p|<|p_{\mathrm{min}}|$. For $t=20$, a longer tail extending to far beyond the core region and interference fringes are present in all cases. For these longer times, the agreement between the HK propagator computation and the TDSE worsens. For the standard CCS, this is also the case, as, in this region, anharmonicities associated with the binding potential become more important. In order to overcome this problem, it is necessary to perform a periodic reprojection of the trajectories onto a static grid. A detailed discussion of this method in the strong-field context will be given elsewhere \cite{Wu_2014} (see also Refs.~\cite{Shalashilin_Jackson_2000,Burant_Batista_2002} for details).

We have also verified that, while the trajectories employed in the HK propagator never cross the separatrix, those used in the CCS method do.
This is a consequence of the fact that the ordered Hamiltonian $H_{\mathrm{ord}}(z^*,z)$ is different from the
classical Hamiltonian $H_{\mathrm{cl}}(p,q)$.  
In fact, as a function of the phase-space coordinates $(q,p)$, for a static field $H_{\mathrm{ord}}(z^*,z)$  is given by
\begin{equation}
 H^{\mathrm{st}}_{\mathrm{ord}}(p,q)=\frac{\gamma} {4}+\frac{p^2}{2}-\left(\frac{\gamma} {\gamma+\lambda} \right)^{1/2}\exp[-\eta q^2]+q\mathcal{E},
\end{equation}
with $\eta=( \lambda\gamma/(\gamma+\lambda) )$.  If the separatrix is however defined with regard to $H_{\mathrm{ord}}(p,q)$ instead of the classical Hamiltonian, the trajectories defined by the CCS method will not cross. In other words, averaging the potential over a Gaussian CS basis leads to the lowering of the barrier, which partially takes tunneling into account. A direct comparison of $H^{\mathrm{st}}_{\mathrm{ord}}(p,q)$ and $H^{\mathrm{st}}_{\mathrm{cl}}(p,q)$ shows an effective energy shift $\gamma/4$ and a shift
\begin{equation}
\Delta V_G(q)=V_G(q)\left[\left(\frac{\gamma} {\gamma+\lambda} \right)^{1/2}\exp\bigg\{\frac{\lambda^2 q^2}{\gamma +\lambda}\bigg\}-1\right]
\end{equation}
in the binding potential, where $V_G(q)$ is given by Eq.~(\ref{eq:gaussian}). For discussions of these shifts see \cite{Miller_2002,Child_2003}.

\section{Approximate estimates}
\label{estimates}
It is a well known fact that, for an inverted harmonic oscillator, IVRs lead to exact descriptions of the time-dependent wave packet dynamics and Wigner quasiprobability distributions exhibit nonlocal behavior \cite{Balazs_1990}. In \cite{Maitra_1997}, however, it has been argued that this does not hold for a general barrier, unless transmission occurs close enough to its top. In this case, the potential barrier may be approximated by an inverted harmonic oscillator, and IVRs give reasonable, albeit not exact, results. It is thus our objective to assess whether, for the potential barriers employed here, the results obtained in the previous section may be justified in this way.

\begin{figure}
\centering
\mbox{\subfigure{\includegraphics[width=6cm]{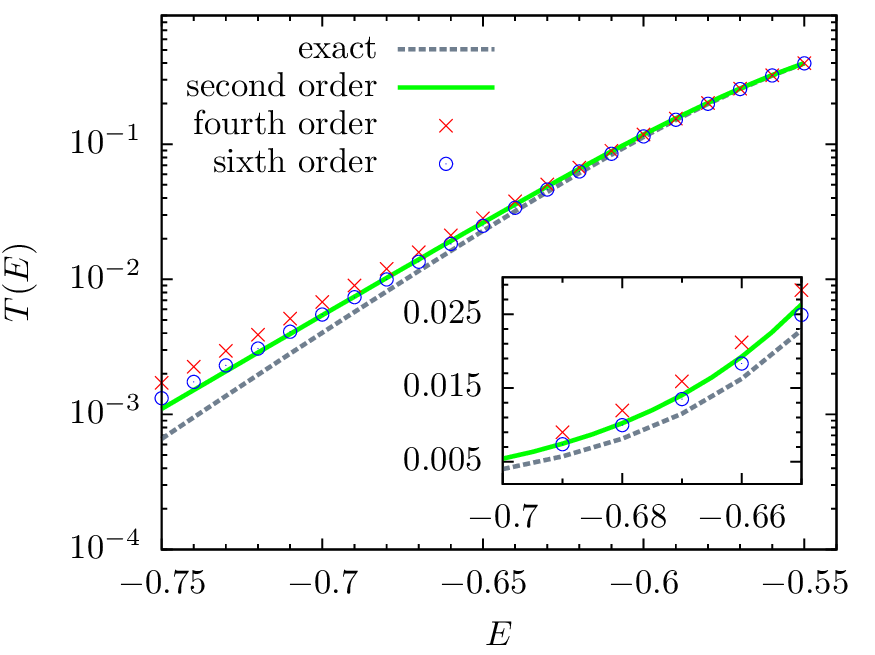}}\quad
\subfigure{\includegraphics[width=6cm]{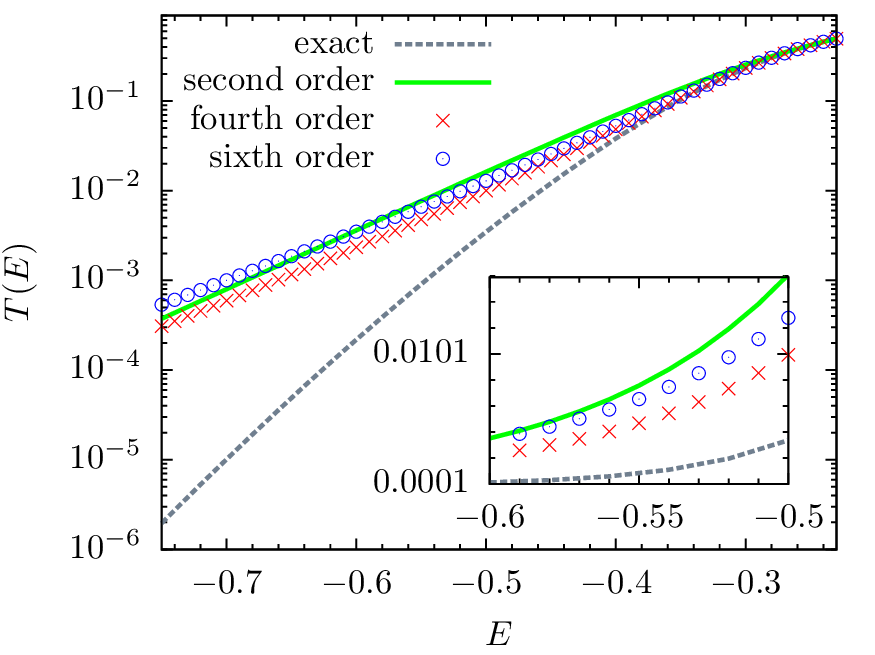}}}
\caption{Transmission coefficients for different order approximations to the barrier in the softcore (left) and the inverted Gaussian potentials (right). The insets show the behavior of such coefficients within the range in which the energy the respective ground state lie.}
\label{fig:transmission}
\end{figure}

For that reason, we expand $V_{\mathrm{eff}}(x)$ around the saddle $x_s$, and, using the uniform WKB approximation, we compute the transmission coefficient through this barrier. This coefficient reads as
\begin{equation}
P(E)=\left(1+\exp\bigg\{2\int_{x_l}^{x_r}\sqrt{2(V_{\mathrm{eff}}(x)-E)}\mathrm{d}x\bigg\}\right)^{-1},
\end{equation}
where $x_l$ and $x_r$ are the left and right turning points, respectively, for which $V_{\mathrm{eff}}(x_l)=V_{\mathrm{eff}}(x_r)=E$. In Fig.~\ref{fig:transmission}, these results are displayed as a function of the energy of the initial wave packet. The figure confirms that the inverted harmonic oscillator is a reasonable approximation, for the parameter range employed in this work. This approximation becomes increasingly more accurate as the energy approaches the threshold and over-the-barrier regime. A better agreement is observed for the soft-core potential. This is consistent with the results in the previous section (see Figs.~\ref{fig:wf1} and \ref{fig:wignergaussian})

\section{High-harmonic generation}
\label{HHG}
We will now establish a connection with previous work in the literature, and  compute high-harmonic spectra with the
HK propagator. We will use the soft-core potential
(\ref{eq:softcore}) and the interaction Hamiltonian (\ref{eq:laserpot}),
but assume that the electronic wave packet is initially localized at the core.

The HHG spectra are given by
\begin{equation}
\sigma(\omega)=\left|\int \mathrm{d}t d(t)\exp(\mathrm{i}\omega t)\right|^2,
\label{eq:spectra}
\end{equation}
where
\begin{equation}
d(t)=-\int^{\infty}_{-\infty}dx \Psi^*(x,t)\frac{\partial V_a(x)}{\partial x}\Psi(x,t),
\end{equation}
denotes the dipole acceleration. The time-dependent wave function is either
given by the solution of Eq.~(\ref{eq:tdsegeneral}) in coordinate space or by $\Psi_{\mathrm{HK}}(x,t)$.

\begin{figure*}[tbp]
\begin{center}
\begin{minipage}[0.5cm]{3in}
\includegraphics[width=6cm]{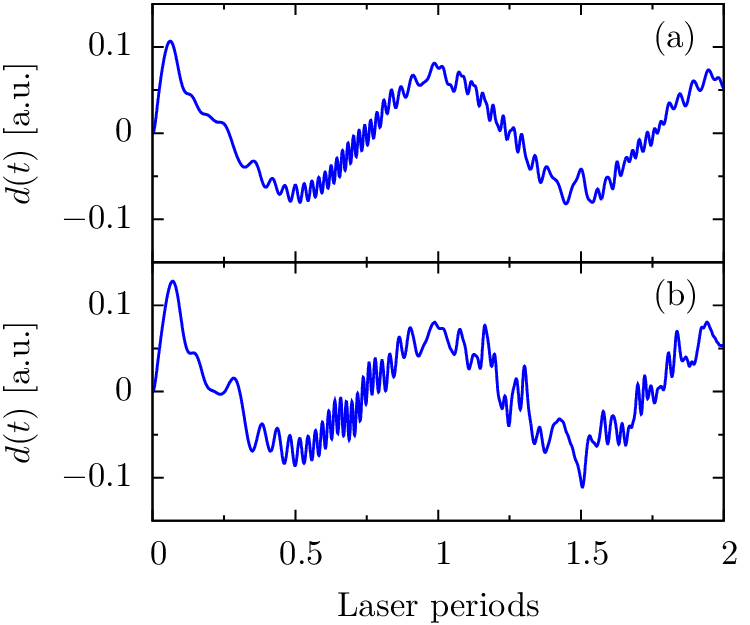}
\end{minipage}
\begin{minipage}[0.5cm]{3in}
\includegraphics[width=6cm]{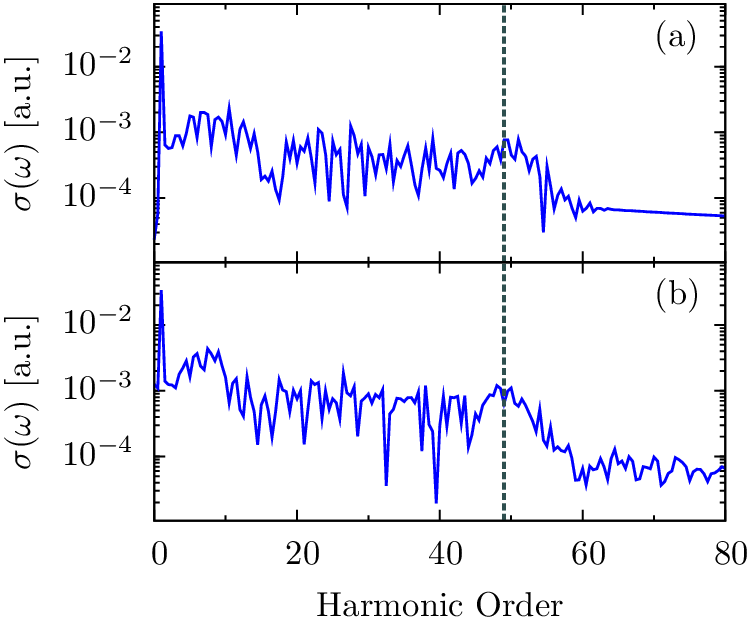}
\end{minipage}
\end{center}
\caption{Dipole acceleration over two field cycles, together with the HHG
spectra (left and right panels, respectively), computed using a laser field
of frequency $\protect\omega=0.05$ a.u. and amplitude $\mathcal{E}=0.075$,
the soft-core potential (\protect\ref{eq:softcore}) and an initial
wavepacket (\protect\ref{eq:initialwp}) of width $\protect\gamma=0.5$,
centered at $q_{\protect\alpha}=0$ and with vanishing momentum $p_{\protect%
\alpha}=0$. Panels (a), (b) and (c) were computed using the full TDSE
computation, the Herman Kluk propagator, and the CCS method, respectively. The dashed
lines show the energy position of the cutoff, which in this case is located
at $I_p+3.17U_p=49\protect\omega$.}
\label{fig:spectra1}
\end{figure*}

In Fig.~\ref{fig:spectra1}, we display $d(t)$, together with its power
spectrum, for an initial wave packet (\ref{eq:initialwp}) centered at $%
q_{\alpha}=0$ and with vanishing momentum $p_{\alpha}=0$, whose width is $%
\gamma=0.5$ a.u. The figure shows a good qualitative agreement between the
fully quantum and semiclassical acceleration, which roughly follow the field
and exhibit a series of high-frequency oscillations. These oscillations,
together with spatial localization, are responsible for the HHG plateau.
They have been identified and discussed in previous publications employing
the TDSE \cite{Protopapas_1996} and the HK propagator \cite%
{Sand_1999,Zagoya_2012,Zagoya_2012_2}, for an initial wave packet far from
the core. They have also been studied in a different context, namely the
adiabatic approximation \cite{Okajima_2012} and Bohmian trajectories \cite{Wu_2013,Wu_2013_2}.
 The agreement between the outcome of the TDSE and the HK
propagator is particularly good for times below half a cycle. Phase
differences however arise between $0.5T$ and $1.5T$, with $T=2\pi/\omega_0$.

This good agreement persists for the spectra, which exhibit a long plateau
followed by a sharp cutoff at $I_p+3.17U_p$ and a reasonably similar
substructure, such as the intensity modulations near the cutoff (harmonic
orders $45\leq  N \leq 59$) and in the below-threshold harmonic region
(harmonic orders $N \leq 19$). Discrepancies, however, exist in the overall
intensity of the plateau, which is slightly higher for the semiclassical
spectrum, and in this substructure. These discrepancies are associated with
the phase differences mentioned above.

Physically, this dephasing is a consequence of the fact that the semiclassical IVRs do not fully account for processes in 
which classically forbidden regions in phase space are crossed, which affect the overall phase of the wave packet. 
Examples of such processes are tunneling ionization and over-the-barrier reflections. We have indeed found that this 
dephasing decreases if the energy of the initial wavepacket is increased, for instance, by changing its width or 
initial momentum $p_{\alpha}$. This, together with an overall decrease in the plateau height, 
is consistent with the fact that, for $p_{\alpha} \neq 0$ or $\gamma\ll1$, there will be an enhancement in the over-the barrier pathways. A similar dephasing has also been observed in the context of the transmission of a wave packet through an Eckart barrier \cite{Grossmann_1995}. These issues may constitute a problem for long times \cite{Maitra_1997}. We have observed that our results are reasonably accurate for $t\leq 3.5 T$. In
contrast, the results from the CCS method are once more practically identical to those of the TDSE.
\section{Conclusions}

\label{conclusions}
The results presented in this paper strongly suggest that semiclassical initial-value representations, a concrete example of which is the Herman Kluk propagator, may be employed for describing strong-field wave-packet dynamics, even if this wave packet is initially bound and located within the core region. We have observed a reasonably good agreement with fully quantum mechanical methods, such as the numerical solution of the time-dependent Schr\"{o}dinger equation (TDSE), or the Coupled Coherent States (CCS) representation, for static and time-dependent fields, and different types of binding potentials. This agreement manifests itself in the time evolution of Wigner quasiprobability distributions, and in the computation of time-dependent quantities such as the dipole acceleration.

 A noteworthy feature is the presence of tails in the Wigner distributions, which leave the core region following the separatrices very closely. Depending on the momenta associated with this tail, it may be related to over-the-barrier or tunneling ionization. Similar tails have been identified in the literature, using either a zero-range potential \cite{Czirjak_2000,Czirjak_2013} or the TDSE \cite{Graefe_2012}. Therein, however, focus has been placed on how this tail behaves outside the core region, and on its agreement with classical trajectories as defined by the strong-field recollision model \cite{Corkum_1993}. In this work, we place more emphasis on the position-momentum correlation in the vicinity of the core as evidence for different ionization mechanisms. For the soft-core, long-range potential employed here, the momentum in this tail indicates substantial over-the-barrier ionization, while for the Gaussian, short-range potential, tunnel ionization seems to be dominant. It therefore appears that the
Wigner function is crossing a classically forbidden region.

Thus, one may argue that, while the trajectories in the semiclassical method used in this paper are
classical and will never cross a separatrix, the initial position-momentum spread will provide the wave function with 
access to classically forbidden regions. This probability density, and hence the Wigner function, 
may exhibit nonlocal behavior around a separatrix. This specific behavior has been shown in \cite{Balazs_1990} 
for an inverted harmonic oscillator.

On the other hand, if the classical trajectories that start outside the bound phase-space region are removed, the tail in the Wigner functions disappears. This implies that they are an essential, and fully classical ingredient for reproducing this tail in the context of a semiclassical initial-value representation. Furthermore, a semiclassical approximation would only be exact for a parabolic barrier, while for the potentials employed here it will be only approximate.
The presence of anharmonicity in more realistic barriers, together with the existence of tunneling loop structures in phase space for anharmonic potentials, was in fact employed in \cite{Maitra_1997,Maitra_1997_2} as a criticism to the findings in \cite{Balazs_1990}. These structures become dominant for long times. Nevertheless, any barrier in the vicinity of its maximum may be approximated by an inverted harmonic oscillator. Specifically, our approximate estimates for the transmission coefficient show that this is a reasonable approximation for the parameter range employed in this work. Another noteworthy aspect is that, in \cite{Maitra_1997,Maitra_1997_2}, the potential barrier is flat, i.e., $\lim_{x\rightarrow \infty}V_a(x)=0$, while, for the effective potential $V_{\mathrm{eff}}(x)$ employed in this work, this condition does not hold. In such references, it was repeatedly emphasized that this condition led to the tunneling contributions becoming dominant for  longer times.

One should bear in mind, however, that the trajectories do need to cross classically forbidden regions for the phase of the wave function to build up correctly. If this does not occur, there will be a degradation of this phase for longer times \cite{Maitra_1997,Maitra_1997_2}. Since the dipole acceleration and the HHG spectra are strongly dependent on this phase, only a qualitative agreement with the full quantum mechanical result may be reached if a standard semiclassical IVR is employed. A quantitative agreement would require more sophisticated approaches, possibly along the lines in \cite{Kay_1997,Ankerhold_2002} or by adding higher order correction terms to the HK propagator as in \cite{Zhang_2003,Kay_2006}.
Still, the results in Sec.~\ref{HHG} show that the HK propagator may be quite useful in modeling strong-field phenomena and understanding quantum-interference effects in, for instance, few-cycle laser pulses, for which this critical regime has not been reached.
\section*{Acknowledgements}
We are grateful to A. Fring, A. Olaya-Castro, A. Serafini, J. Guo, X. S. Liu and in particular to F. Grossmann and J. M. Rost for very useful discussions. C.F.M.F., C.Z.M. and J.W. would like to thank the University of Leeds and the Max Planck Institute for Physics of Complex Systems, Dresden, where the final stages of this work were carried out, for their kind hospitality. This work was funded by the UK EPSRC (Grant EP/J019143/1) and the CSC/BIS (China-UK Studentship for Excellence).
\appendix
\section{Relation between the Coupled Coherent States and the Herman Kluk initial value representations}
%\numberwithin{equation}{section}
 \label{app:A}
  For the reader's convenience, in this appendix we provide a brief sketch of the HK and of the CCS representation, and emphasize how they share a common origin. The details of the derivation of CCS working equations and the HK formula can be found in Ref. \cite{Shalashilin_2004}.  Originally, the semiclassical phase-space HK method has been derived differently \cite{Herman_1984}. Nonetheless, the current derivation shows that it also can be obtained as an approximation of the exact quantum dynamics in phase space.
\subsection*{Integro-differential form of the Schr\"odinger equation in the Coherent State representation}
Gaussian coherent States (CS) are eigenstates of the creation and annihilation operators
\begin{equation}\hat{a}|z\rangle=z|z\rangle \quad \mathrm{and} \quad \langle z|\hat{a}^{\dagger}=\langle z|z^*,\end{equation}
where the eigenvalue $z(q,p)$ is a complex number in phase space parametrized as in Eq.~(\ref{eq:CS}).
Coherent States are not orthogonal; their overlap is given as
\begin{equation}\langle z_l|z_j\rangle=\Omega_{lj}=\exp\left(z_l^*z_j-\frac{z_l^*z_l}{2}-\frac{z_j^*z_j}{2}\right).
\label{eq:overlap}\end{equation}
The identity operator in the CS representation reads as
\begin{equation}\hat{I}=\int|z\rangle\langle z| \frac{\mathrm{d}^2z}{\pi},\label{eq:identity}\end{equation}
where $\mathrm{d}^2z=\mathrm{d}q\mathrm{d}p/2$ is the notation for phase space integration.

By inserting the identity operator in the time-dependent Schr\"odinger equation (\ref{eq:tdsegeneral}) and closing it with $\langle z|$, one obtains the  integro-differential equation
\begin{eqnarray}
&\int\langle z|z^\prime \rangle\frac{\mathrm{d}D_{z'}(t)}{\mathrm{d}t}\exp\left[\mathrm{i}\left(S_{z^\prime}-S_{z}\right)\right]\frac{\mathrm{d}^2z'}{\pi}= \nonumber \\
&- \mathrm{i}\int \langle z|z^\prime \rangle
\delta^2H_{\mathrm{ord}}^{'*}(z^*,z^\prime)\exp\left[\mathrm{i}\left(S_{z^\prime}-S_{z}\right)\right]D_{z'}(t)\frac{\mathrm{d}^2z'}{\pi}\label{eq:Sch_D}\end{eqnarray}
for the amplitudes $D_{z^{\prime}}(t)$, where the coupling kernel is
\begin{equation}
\delta^2H_{\mathrm{ord}}^{'*}(z^{*},z^\prime)=H_{\mathrm{ord}}(z^{*},z^\prime)-H_{\mathrm{ord}}(z^{\prime*},z^\prime)-\mathrm{i}\frac{\mathrm{d}z^{\prime}}{\mathrm{d}t}(z^*-z^{\prime
*})\label{eq:deltaH},\end{equation} with the matrix elements of the ordered Hamiltonian defined in the main body of the article (see Eq.~\ref{eq:Hord}). An important property of (\ref{eq:deltaH}) is that the kernel $\delta^2H_{\mathrm{ord}}^{'*}(z^{*},z^\prime)$ is always small for $z$ and $z^\prime $ close to each other and their overlap $\langle z|z^\prime \rangle$ vanishes when they are far away. The exact
form of the kernel depends on the choice of the trajectories, which
are used to guide the basis, and it becomes particularly simple if
they are determined by Hamilton's equations with a quantum
averaged Hamiltonian
\begin{equation}\frac{\mathrm{d}z^{\prime}}{\mathrm{d}t}=-\mathrm{i}\frac{\partial H_{\mathrm{ord}}(z^{\prime *},z')}
{\partial z^{'*}}.\label{eq:hamiltonseq}\end{equation}
In practice, a finite basis of Gaussian CSs is used so that the integral in (\ref{eq:Psi_D}) becomes a finite sum. The identity operator is discretized as
\begin{equation}
\hat{I}=\int|z\rangle\langle z|\frac{\mathrm{d}^2z}{\pi}\approx\sum_{i,j}|z_i\rangle(\Omega^{-1})_{ij}\langle z_j|
\end{equation}
where $\Omega^{-1}$ is the inverse of the overlap matrix $\Omega_{ij}=\langle z_i| z_j\rangle$. 
This specific discretization reduces the integro-differential equation (\ref{eq:Sch_D}) to a system of linear 
equations for the derivates $\dot{D}_{z}(t)$, 
which are solved numerically together with Eq.~(\ref{eq:hamiltonseq}) for the trajectories.

\subsection*{HK propagator and analytical solution of the integro-differential 
form of the Schr\"odinger equation in the local quadratic approximation}

The Herman-Kluk method can also be derived from Eq.~(\ref{eq:Sch_D}) \cite{Shalashilin_2004} 
by employing the local quadratic approximation, which only takes into account the first and second terms in the 
Taylor expansion of the potential energy around a specific trajectory. 
This allows to use classical trajectories instead of those driven by the quantum averaged 
Hamiltonian $H_{\mathrm{ord}}(z^{\prime *},z')$. Furthermore, the local quadratic approximation assumes that ony the CSs which are very close to each other, and therefore are driven by the same quadratic potential, are coupled. Under these assumptions,
the integrals in (\ref{eq:Sch_D}) may be calculated analytically. 

The resulting wave function may be then represented as
\begin{equation}
|\Psi_{\mathrm{HK}}(t)\rangle=\int |z\rangle\sqrt{M_{zz}}\mathrm{e}^{\mathrm{i}S^{\mathrm{cl}}_z}\langle z_0|\Psi(0)\rangle\frac{\mathrm{d}^2z_0}{\pi}, \label{eq:HK_prop}
\end{equation}
where $\sqrt{M_{zz}}$ denotes the HK prefactor in the $z$ notation. In this notation 
$M_{zz}$ is a single element of the monodromy (or stability) matrix
\begin{equation}
{\bf M}=
\left(\begin{array}{cc}
M_{zz} & M_{zz^*} \\
M_{z^*z} & M_{z^*z^*}  \\
\end{array} \right)\, ,
\end{equation}
which describes how stable the dynamics around a specific trajectory are. The monodromy matrix elements in this representation are related to those in the $p,q$ representation by
\begin{eqnarray}\nonumber
M_{zz}&=&2^{-1}(m_{qq}+m_{pp}-\mathrm{i}\gamma m_{qp}+\mathrm{i}\gamma^{-1}m_{pq}), \\ \nonumber
M_{zz^*}&=&2^{-1}(m_{qq}-m_{pp}+\mathrm{i}\gamma m_{qp}+\mathrm{i}\gamma^{-1}m_{pq}), \\ \nonumber
M_{z^*z}&=&2^{-1}(m_{qq}-m_{pp}-\mathrm{i}\gamma m_{qp}-\mathrm{i}\gamma^{-1}m_{pq}), \\
M_{z^*z^*}&=&2^{-1}(m_{qq}+m_{pp}+\mathrm{i}\gamma m_{qp}-\mathrm{i}\gamma^{-1}m_{pq}).
\end{eqnarray}

The evolution of the monodromy matrix is given by the matrix of second derivatives of the Hamiltonian
\begin{equation}
\frac{\mathrm{d} \mathbf{M}}{\mathrm{d}t}=
\left(\begin{array}{cc}
\partial^2H_{cl}/\partial z^*\partial z & \partial^2H_{cl}/\partial {z^*}^2  \\ \\
 -\partial^2H_{cl}/\partial {z}^2  & -\partial^2H_{cl}/\partial z^*\partial z  \\
\end{array} \right)
\left(\begin{array}{cc}
M_{zz} & M_{zz^*} \\
M_{z^*z} & M_{z^*z^*}  \\
\end{array} \right).
\end{equation}

In this work only a 1D case has been considered. However, a 
generalization to more than one degree of freedom is
straightforward, as a multidimensional CS is simply a product of 1D CSs.
The CCS method can also be generalized for multielectronic states by introducing Fermionic Coherent States
with proper permutational symmetry \cite{Kirrander_2011}. 
\section*{References}
\bibliographystyle{unsrt}
\bibliography{ReviewCarla_final}
{}

\end{document}